\documentclass[11pt]{article}
\usepackage{tikz}
\usepackage{geometry}
\usepackage[english]{babel}
\usepackage[utf8]{inputenc}
\usepackage[T1]{fontenc}
\usepackage{indentfirst}
\usepackage{amsmath}
\usepackage{amssymb}
\usepackage{amsthm}
\usepackage{proof}
\usepackage{eufrak}
\usepackage{graphicx}
\usepackage{psfrag}
\usepackage{physics}

\usepackage[nosort]{cite}
\usepackage{subfigure}
\usepackage{graphicx}
\usepackage{xcolor} 
\usepackage{bbm}

\allowhyphens

\pgfdeclarelayer{edgelayer}
\pgfdeclarelayer{nodelayer}
\pgfsetlayers{edgelayer,nodelayer,main}
\tikzstyle{none}=[inner sep=0pt]
\tikzstyle{rn}=[circle,fill=Red,draw=Black,line width=0.8 pt]
\tikzstyle{gn}=[circle,fill=Lime,draw=Black,line width=0.8 pt]
\tikzstyle{yn}=[circle,fill=Yellow,draw=Black,line width=0.8 pt]
\tikzstyle{simple}=[-,draw=Black,line width=2.000]
\tikzstyle{arrow}=[-,draw=Black,postaction={decorate},decoration={markings,mark=at position .5 with {\arrow{>}}},line width=2.000]
\tikzstyle{tick}=[-,draw=Black,postaction={decorate},decoration={markings,mark=at position .5 with {\draw (0,-0.1) -- (0,0.1);}},line width=2.000]
\usetikzlibrary{arrows}
\tikzstyle{black}=[fill=black, shape=circle]
\tikzstyle{arrow}=[->]
\tikzstyle{red arrow}=[draw=red, ->]


\newcommand{\be}{\begin{equation}}
	\newcommand{\ee}{\end{equation}}
\newcommand{\bea}{\begin{eqnarray}}
	\newcommand{\eea}{\end{eqnarray}}

\newcommand{\q}{\mathbb{Q}}
\newcommand{\h}{\mathcal{H}}
\newcommand{\g}{\mathcal{G}}
\newcommand{\ct}{\otimes}
\newcommand{\eps}{\epsilon}
\newcommand{\s}{\sigma}
\newcommand{\pa}{\partial}

\newcommand{\unit}{\mathbbm{1}}

\usepackage{ifpdf}

\ifpdf
\usepackage{epstopdf}
\usepackage[pdftex,colorlinks,urlcolor=blue,citecolor=blue,linkcolor=black]{hyperref}
\else
\usepackage[hypertex,colorlinks,urlcolor=blue,citecolor=blue,linkcolor=black]{hyperref}
\fi
\newcommand{\nn}{\nonumber}
\newcommand{\ysla}[1]{#1}
\begin{document}
	
	\thispagestyle{empty}

	\begin{center}%
		{\LARGE\textbf{\mathversion{bold}%
				On deforming and breaking integrability}\par}
		
		\vspace{1cm}
		{\textsc{Ysla F. Adans$^{a,b}$, Marius de Leeuw$^{a}$, Tristan McLoughlin$^{a}$ }}
		\vspace{8mm} \\
		\textit{
			$^a$ School of Mathematics \& Hamilton Mathematics Institute, \\
			Trinity College Dublin, Ireland\\
			[5pt]
		}
		\vspace{.2cm}
		\textit{
			$^b$   Universidade Estadual Paulista (Unesp), Instituto de Física Teórica (IFT), São Paulo, Brasil. \\
			[5pt]
		}
		
		\texttt{ysla.franca@unesp.br}, \texttt{deleeuwm@tcd.ie}, 
		\texttt{tristan@maths.tcd.ie} \\
		%

		%%%%%%%%
		\par\vspace{15mm}

		\textbf{Abstract} \vspace{5mm}
		
		\begin{minipage}{13cm}
			In this paper we study nearest-neighbour deformations of integrable models. After expanding in the deformation parameter, we identify four possible types of deformations. First there are deformations that simply break or preserve integrability. Then we find two different subtle cases. The first case is where the deformation is only integrable if all orders of the deformation parameter are taken into account. An example of these are the long-range deformations that appear in holographic models. The second case is when the deformation is perturbatively integrable to some order in the deformation parameter but can not be extended to an integrable model. In this paper we work this out for the XXZ spin chain and discuss the level statistics of each of these cases. We find numerical evidence that the onset of chaos occurs differently in each of these models. For the perturbatively integrable models, we find that the deformation strength at which chaos appears demonstrates a volume-scaling intermediate between strong and weak integrability breaking models. 
		\end{minipage}
	\end{center}
	
	%%%%%%%%%%%
	\newpage 
	
	\tableofcontents
	\bigskip
	\hrule
	
	\section{Introduction}
	
	Due to their large amount of symmetry, integrable models usually exhibit special and useful features such as solvability. Generically, perturbing an integrable system will break integrability and will spoil these features to some degree. Actually, there is a more refined spectrum of possibilities when an integrable model is perturbed. It turns out that integrability can be broken to various degrees and in this paper we set out to describe different scenarios and study the effect of the integrability breaking on the spectral statistics.
	
	A lot of progress was recently made in the classification of integrable models due to the development of the boost operator method \cite{DeLeeuw:2019gxe, deLeeuw:2020ahe , deLeeuw:2020xrw, Corcoran:2023zax}. The key idea is to use the boost operator to generate a tower of conserved charges and impose that they commute. In this way it is possible to classify solutions of the quantum Yang--Baxter equation. It was used to study integrable deformations of scattering matrices that for example appear in the AdS/CFT correspondence \cite{ deLeeuw:2020ahe, deLeeuw:2021ufg}. However, this approach to studying deformations is rather indirect as it uses the full classification and then finds in which general solutions of the Yang-Baxter equation a given model is contained.
	
	In this paper we take a more direct approach to studying deformations of an integrable model by looking at the level of the charges using the boost operator formalism. We will use an integrable Hamiltonian $H^{(0)} = H_{\text{int}}$ as a starting point and then add a perturbation
	\begin{align}
		H = H^{(0)} + \eps H^{(1)} + \ldots
	\end{align}
	and put this in the boost operator formalism. A similar approach was already applied to long-range deformations in \cite{deLeeuw:2022ath}, see also \cite{Gombor:2021nhn}. For long-range deformations a framework on the level of the charges was developed in \cite{Bargheer:2008jt,Bargheer:2009xy}. The key ingredient is the so-called deformation equation with which deformations with increasing interaction range can be generated. Integrable deformations using the deformation equation formalism for the XXZ spin chain were studied in \cite{beisert_integrable_2013} and the corresponding long-range deformations were classified and found to be in agreement with the general framework. In \cite{deLeeuw:2022ath} it was shown that the boost approach was able to reproduce all of these deformations. 
	
	We identify four possible types of deformations. First there are the deformations that simply break integrability. This is of course the generic situation where you add some random matrix $\h^{(1)} $ as perturbation. Second, there are models for which the perturbation is exactly integrable. An example of this is the XXX spin chain to which you add a $\sigma_z \otimes \sigma_z$ term, which turns it into the XXZ spin chain.  The third and fourth cases are more interesting. The third case is where the deformation is only integrable if all orders of the deformation parameter are taken into account. An example of these are the long-range deformations that appear in holographic models such as studied in  \cite{Bargheer:2008jt,Bargheer:2009xy}. The fourth case is when the deformation is integrable to some order in the deformation parameter but can not be extended to an integrable model beyond that. An example of such a model is derived in this paper and takes the form
	\begin{align}
		H_{\text{QInt}}=H_{\text{XXZ}}+\epsilon \alpha \sum_{i=1}^L (\sigma_{x,i} \sigma_{x,i+1}-\sigma_{y,i} \sigma_{y,i+1})+\epsilon \beta \sum_{i=1}^L \sigma_{z, i}+\epsilon \gamma  \sum_{i=1}^L (\sigma_{x,i}\sigma_{y,i+1}-\sigma_{y, i}\sigma_{x, i+1}).
	\end{align}
	This is a deformation of the  XXZ spin-chain for $\epsilon=0$ but is only integrable to $\mathcal{O}(\epsilon)$ for non-zero coefficients. It is interesting to note that each of the deformations separately preserve integrability, for example $\beta=\gamma=0$ is simply the XYZ Hamiltonian. The second deformation corresponds to the $z$-component of the total magnetisation, $\sigma^{\text{tot}}_z=\sum_{i=1}^L \sigma_{z,i}$, while the third corresponds to a twist in the boundary conditions. However, for $H_{\text{XYZ}}$ rotations about the $z$-axis are no longer a symmetry and this removes the possibility of exchanging modified local interactions for twisted boundary conditions. More precisely, the term proportional to $\gamma$ is called the anti-symmetric or Dzyaloshinskii–Moriya interaction (DMI) term; see \cite{camley2023consequences} for a review. It results in a spiral structure that has been associated with magnetic skyrmions. In practical applications where such an interaction appears, the DMI strength is just $1\%$ of the ferromagnetic coupling, which fits perfectly with our perturbative approach. This interaction and its spectrum have been studied by different approaches \cite{PERK1976115, alcaraz1990heisenberg, beisert_integrable_2013, feher2019propagator}.
	
	The breaking of integrability, the onset of quantum chaos and eigenstate thermalisation are deeply intertwined concepts which have become increasingly relevant to experiments e.g. \cite{kinoshita2006quantum, wei2022quantum, scheie2021detection, rosenberg2024dynamics, Chen:2026xft}.
	Generic many-body systems that demonstrate quantum chaos are expected to thermalise: when evolved from out-of-equilibrium states, local observables relax to values consistent with the predictions of statistical mechanics and can be characterized by a relatively small set of variables. In contrast, integrable systems possess an extensive number of conserved charges which constrain their dynamics. They do not thermalise in the standard sense but rather relax to states that can be described by the Generalised Gibbs Ensemble (GGE) \cite{rigol2007relaxation,Gogolin:2015gts, essler2016quench}. 
	
	The intermediate regime between integrability and full quantum chaos remains less well understood. In the thermodynamic limit, it is expected that the onset of chaos will be sharp with non-chaotic behaviour appearing only in exactly integrable models. There is evidence that in systems which are close to integrable, the evolution from out-of-equilibrium states involves an initial pre-thermalisation phase \cite{moeckel2008interaction, rosch2008metastable, kollar2011generalized, bertini2015prethermalization} where the integrable dynamics dominate and observables reach a quasi-stationary state described by a GGE before reaching thermal equilibrium on much longer time-scales $\tau \propto \epsilon^{-2}$. This scaling is not completely universal and there are models that thermalise on even longer time scales e.g. \cite{jung2006transport, Durnin:2020kcg} and it has been suggested that this is related to the presence of quasi-conserved charges \cite{PhysRevB.105.104302, Orlov_2023,PhysRevResearch.5.043019, Vanovac:2024tkj}.
	
	In finite dimensional systems there is a distinct intermediate region where, as one deforms the integrable model, standard measures of chaos such as level statistics become progressively more chaotic \cite{Rabson, santos2010onset, rigol2010quantum}. In section \ref{sec:stats} we see that this is true for the class of models we consider and also that there is some variation in exactly how the models behave in the intermediate region. The system-size dependence of the onset of chaos in this region has a non-trivial structure that is only partly understood. Numerical evidence \cite{modak2014universal, modak2014finite} suggests that for gapless systems the critical coupling, $\epsilon_c$, at which chaos develops scales as $\sim L^{-3}$. The authors  \cite{modak2014universal, modak2014finite} demonstrated that this result was robust against variations in the integrability-breaking perturbation and conjectured that this scaling was universal for gapless systems. For systems with a gap they proposed  a faster-than-power-law dependence on the system size. A more detailed description of the onset of chaos was developed in \cite{bulchandani2021onset} using an analogy with delocalisation in quantum dots \cite{altshuler1997quasiparticle}.  In this framework, a distinction was made between  \textit{transitions}, generated by perturbations which correspond to short-range interactions in Fock space and produce a scaling $1/(L^{(n_h+1)/2}\ln L)$ where $n_h$ is the Hamming distance, and \textit{crossovers}, generated by perturbations which are long-range in Fock space producing exponential scaling of the critical coupling. While this picture provides a compelling view of integrability breaking, for strongly interacting systems the Fock space picture is difficult to access analytically. 
	
	The long-range integrable spin-chains studied in the context of the AdS/CFT correspondence provide a particularly interesting class of models in this context. Truncating such models at order $\epsilon$ results in a perturbed integrable model with quasi-conserved charges which commute with the Hamiltonian up to corrections at order $\epsilon^2$. These models can thus be considered as breaking integrability in a ``weak" sense\footnote{The terminology ``weak-breaking of integrability" is often used to refer to the generic case where the integrability-breaking coupling is small. Here it denotes the specific algebraic structure of the deformation term which has also been called ``nearly-integrable" in the literature.}.  These models are additionally of interest as they can, at least in part, also be viewed as originating from $T\bar{T}$ deformations \cite{Pozsgay:2019ekd, Marchetto:2019yyt} and are exactly the class of models which are expected to show anomalously slow thermalisation \cite{PhysRevB.105.104302, PhysRevResearch.5.043019}. In \cite{Szasz-Schagrin:2021pqg} it was shown that this weak breaking is reflected in spectral statistics and in the scaling of the critical coupling. For a current deformation of the XXZ model in the gapless phase, the onset of chaos scales as $\epsilon_c \sim L^{-2}$, while in the gapped phase an exponential-in-$L$ scaling was observed which was nonetheless weaker than that induced by a generic next-to-nearest-neighbour deformation.  A similar weak scaling, in this case $\epsilon_c \sim L^{-1}$, was found in \cite{McLoughlin:2022jyt} for the $\mathcal{N}=4$ super-Yang–Mills spin chain in the $\mathfrak{su}(2)$ sector truncated at two loops. This model is equivalent to the $\mathfrak{su}(2)$-invariant next-to-nearest-neighbour deformation of the XXX Heisenberg chain.
	The model $H_{\text{QInt}}$ is especially notable in this context. It possesses quasi-conserved charges but it cannot be made integrable by the addition of further terms and so lies between the generic case and the weak-breaking models. In section \ref{sec:stats} we present numerical evidence that this intermediate character is reflected in the volume dependence of its critical coupling which can be described by a power-like expression with an intermediate exponent $\epsilon_c \sim L^{-2.5}$.
	
	The paper is organised as follows: in section \ref{sec:defs} we review the boost method for the perturbative construction of deformed conserved charges in local integrable models and in section \ref{sec:XXZ_def} we apply this to all nearest-neighbour deformations of the XXZ spin-chain. Given the Hamiltonian, the R-matrix can be obtained from the Sutherland equation and this is done in section \ref{sec:Rmat}. Section \ref{sec:stats} considers the spectral statistics of the deformed XXZ models, particularly $H_{\text{QInt}}$, and studies the volume dependence of the critical coupling. Additional calculational details are provided in the appendices \ref{app.g}, \ref{app:data}.

	\section{Integrable Deformations using the Boost Operator}
	\label{sec:defs}
	
	\paragraph{Deformations}
	Consider an integrable Hamiltonian density with nearest-neighbour interactions $\h_{ij}^{(0)}$. We will consider deformations of this Hamiltonian 
	\begin{align}
		\h_{ij}^{(0)} \rightarrow \h_{ij}(\eps) = \h_{ij}^{(0)}  + \eps \h_{ij}^{(1)}  + \eps^2 \h_{ij}^{(2)}  + \ldots, \qquad j=i+1,
	\end{align}
	described by the deformation parameter $\eps$.  If we then consider a spin chain with periodic boundary conditions,  we obtain a deformation of our spin chain Hamiltonian $\q_2$
	\begin{equation} \label{q2.int}
		\q_2(\eps)= \sum_{i=1}^{L} \h_{i,i+1}(\eps) = \q_2^{(0)}+ \eps \q_2^{(1)} +\ldots
	\end{equation}
	We will now study which of these deformations break or preserve integrability by using the boost operator formalism.

	\paragraph{Boost formalism}
	In order to apply the boost operator formalism, the first step is to calculate the next conserved charge $\q_3$ in this new configuration. We will follow \cite{deLeeuw:2022ath} and consider $\q_3$ given by the boost operator
	\begin{equation} \label{q3.t}
		\q_3(\eps)  = \sum_{i=1}^{L} \comm{\h_{i,i+1}(\eps)}{\h_{i-1,i}(\eps)} + \g_{i,i+1}(\eps)
	\end{equation}
	where $\g$ is a range two operator which also allows for a power series expansion in $\eps$. We then impose the necessary condition for integrability 
	\begin{align}
		\comm{\q_2(\eps)}{\q_3(\eps)}=0
	\end{align}
	order by order in $\eps$. To lowest order this equation is satisfied by our assumption that $\h^{(0)}$ is integrable, \textit{i.e.} there is a $\g^{(0)}$ such that 
	\begin{align} \label{int.0}
		\comm{\q_2^{(0)}}{\q_3^{(0)}}=0
	\end{align}
	To first order we obtain a set of linear equations for $\h^{(1)}$ and $\g^{(1)}$:
		\begin{align} \label{int.1}
		\comm{	\q_2^{(0)}}{\q_3^{(1)}}+\comm{	\q_2^{(1)}}{\q_3^{(0)}}=0.
	\end{align}
	 For a given solution of the first order equations, we find for the second order
	\begin{align} \label{int.2}
		\underbrace{	\comm{	\q_2^{(0)}}{\q_3^{(2)}}}_{\text{quadratic in $h_{ij}^{(1)}$, linear in $h_{ij}^{(2)}, g_{ij}^{(2)}$ }}+\underbrace{\comm{	\q_2^{(2)}}{\q_3^{(0)}}}_{\text{linear in $h_{ij}^{(2)}$}}+\underbrace{\comm{	\q_2^{(1)}}{\q_3^{(1)}}}_{\text{quadratic in $h_{ij}^{(1)}$}}=0.
	\end{align}
	which leads to linear equations in $\h^{(2)}$ and $\g^{(2)}$ but also to additional quadratic constraints for $\h_{ij}^{(1)}$, \ysla{Indeed, because the boost-generated charge $\q_3$ is quadratic in the Hamiltonian density, first-order deformations of $\h$ generate second-order terms in higher conserved charges. This yields second-order equations that still constrain the first-order deformations.} In this way we will find models that are integrable to, for example, first order in $\eps$ but will break integrability one order higher. 
	
	One can continue in the same manner and add higher orders. Although we do not have a proof of this, we observe that the third- and higher-order equations do not impose new constraints on the first-order deformations $\h_{ij}^{(1)}$. We find that at any given order is affected at most by the equations one order higher. \ysla{Nonetheless, we caution that this behavior might be coincidental or a direct consequence of the Hilbert space dimension.}

	\section{Deformations of the XXZ spin chain}
	\label{sec:XXZ_def}
	\paragraph{Parameterization}
	In order to demonstrate our formalism we will consider the XXZ spin chain  with Hamiltonian density given by
	\begin{align}
		%\h^{\text{XXZ}} &= \frac{1}{2} \left[\s_x \ct \s_x +\s_y \ct \s_y +(1+\eta)\left(\s_z \ct \s_z \right) + \left(1-\eta\right) \; \unit_4 \right] \\
		\h^{(0)}  =\h^{\text{XXZ}} &=
		\left(
		\begin{array}{cccc}
			\coth (\eta ) & 0 & 0 & 0 \\
			0 & 0 & \text{csch}(\eta ) & 0 \\
			0 & \text{csch}(\eta ) & 0 & 0 \\
			0 & 0 & 0 & \coth (\eta ) \\
		\end{array}
		\right) \\
		& = \frac{\text{csch}(\eta)}{2} \Big[ \sigma_x \otimes \sigma_x + \sigma_y \otimes \sigma_y +\cosh(\eta) (1+ \sigma_z \otimes \sigma_z) \Big]. \label{HXXZ}
	\end{align}
	where $\s_i$ are the Pauli Matrices. It turns out that this model has a more interesting structure than the XXX spin chain ($\eta=0$). In the case of the XXX spin chain, we find that \textit{any} deformation is integrable at first order. At second order we do find constraints, but the equations \ysla{have numerous branches that lead to a zoo of models, the full classification of which lies beyond the scope of this work}. Hence in what follows we will always restrict to the case where $\eta$ is a generic non-vanishing real number. \ysla{Accordingly, our analysis covers both gapped and gapless regimes regimes, but  excludes isolated points such as the XX model, which exhibits higher symmetry and displays behavior closer to the XXX model.}

	We consider a general deformation of the following form
	\begin{equation}
		\h^{(i)}= \; \h_{a}^{(i)}\;+\h_{b}^{(i)}
	\end{equation}
	where
	\begin{equation} \label{def.triv}
		\begin{split}
			\h_{a}^{(i)}= &\; h_0^{(i)}\; \unit \ct \unit +\s_{xy}^{(i)}\; \left(\s_x \ct \s_x +\s_y \ct \s_y\right)+\Delta_{xy}^{(i)}\; \left(\s_x \ct \s_x -\s_y \ct \s_y\right)+\Delta_{z}^{(i)}\;\s_z \ct \s_z\\
			&+\delta_{x,-}^{(i)}\;\left(\s_x \ct \unit-\unit\ct \s_x \right)+\delta_{y,-}^{(i)}\;\left(\s_y \ct \unit-\unit\ct \s_y \right)+\delta_{z,-}^{(i)}\;\left(\s_z \ct \unit-\unit\ct \s_z \right)
		\end{split}
	\end{equation}
	and
	\begin{equation}\label{def.ntriv}
		\begin{split}
			\h_{b}^{(i)}&= \; \delta_{x,+}^{(i)}\;\left(\s_x \ct \unit+\unit\ct \s_x \right)+\delta_{y,+}^{(i)}\;\left(\s_y \ct \unit+\unit\ct \s_y \right)+\delta_{z,+}^{(i)}\;\left(\s_z \ct \unit+\unit\ct \s_z \right)\\
			&+h_{xz,+}^{(i)}\;\left(\s_x \ct \s_z + \s_z \ct \s_x\right)+h_{xz,-}^{(i)}\;\left(\s_x \ct \s_z - \s_z \ct \s_x\right)+h_{yz,-}^{(i)}\;\left(\s_y \ct \s_z - \s_z \ct \s_y\right)\\
			&+h_{yz,+}^{(i)}\;\left(\s_y \ct \s_z + \s_z \ct \s_y\right)+h_{xy,+}^{(i)}\;\left(\s_x \ct \s_y + \s_y \ct \s_x\right)+h_{xy,-}^{(i)}\;\left(\s_x \ct \s_y - \s_y \ct \s_x\right).
		\end{split}
	\end{equation}
	In this notation, the upper index $i$ indicates the order of deformation. We also write $\g^{(i)}$ using the same basis, adding a $g$ to the coefficients, i.e, $h_0^{(i)} \to gh_0^{(i)} $.
	
	\paragraph{First order}
	Notice that all the coefficients from $ \h_{a}^{(1)}$  will preserve integrability as its elements are integrable by construction. It contains the identity deformation, three terms representing XYZ models (which are known to be integrable), and three terms that vanish on periodic spin chains. These latter terms typically correspond to local basis transformations \cite{deLeeuw:2020xrw}.
	
	Imposing the integrability condition at first order, one can find a set of restrictions upon the coefficients of the first order deformation given by:
	\begin{equation} \label{int.o1.1}
		h_{xz,-}^{(1)}=h_{yz,-}^{(1)}=0, \qquad \delta_{x,+}^{(1)}=\delta_{y,+}^{(1)}=0,
	\end{equation}
	together with another set for $g^{(1)}$
	\begin{subequations} \label{g.rel}
		\begin{align} 
			&g\Delta_{xy}^{(1)}=gh_{xy,+}^{(1)}=gh_{xz,-}^{(1)}=gh_{yz,-}^{(1)}=g\delta_{x,+}^{(1)}=g\delta_{y,+}^{(1)}=0\\[0.25cm]
			&g\Delta_{z}^{(1)}=\text{csch}(\eta )\;g\s_{xy}^{(1)}, \quad gh_{xz,+}^{(1)}= -i\; \text{tanh}\left(\frac{\eta }{2}\right)\; \delta_{y,-}^{(1)},\\
			& gh_{xy,-}=-2 i \;\text{csch}(\eta )\;\delta_{z,+}^{(1)},\quad gh_{yz,+}^{(1)}= i\; \text{tanh}\left(\frac{\eta }{2}\right)\; \delta_{x,-}^{(1)}.
		\end{align}
	\end{subequations}
	This means that besides the 7 trivial terms given by \eqref{def.triv}, we find 5 further deformations that preserve integrability up to first order:
	\begin{itemize}
		\item[$\star$] 1 sum of the spins in the $z$ direction: $\s_z \ct \unit+ \unit \ct \s_z$;
		\item[$\star$] 3 sums of mixed spin terms in all directions: $\s_i \ct \s_j+\s_j \ct \s_i$;
		\item[$\star$] 1 difference of mixed spin terms in the $xy$ direction: $\s_x \ct \s_y-\s_y \ct \s_x$~;
	\end{itemize}
	This leads to the following Hamiltonian which is integrable to first order \footnote{\ysla{ $\h_{\text{XYZ}}$ appears since we do not fix specific values for $\sigma_{xy}$ and $\Delta_{xy}$.}}
	\begin{align} \label{def.1st}
		\h_{\text{Int}}^{(1)}= &\; \h_{\text{XYZ}}^{(1)}+\delta_{z,+}^{(1)}\;\left(\unit\ct \s_z +\s_z \ct \unit\right)+h_{xz,+}^{(1)}\;\left(\s_x \ct \s_z + \s_z \ct \s_x\right)\\
		&+h_{yz,+}^{(1)}\;\left(\s_y \ct \s_z + \s_z \ct \s_y\right)+h_{xy,+}^{(1)}\;\left(\s_x \ct \s_y + \s_y \ct \s_x\right)+h_{xy,-}^{(1)}\;\left(\s_x \ct \s_y - \s_y \ct \s_x\right).\nonumber
	\end{align}
	For clarity and conciseness, we suppress the coefficients that do not have a physical effect such as the identity and the terms that vanish on periodic chains. 
	
	Conversely, turning on any of the other 4 parameters will break integrability
	\begin{itemize}
		\item[$\star$] 2 sums of the spins in the $x,y$ directions: $\s_i \ct \unit+ \unit \ct \s_i, \quad i=x,y$~;
		\item[$\star$] 2 differences of mixed spin terms in the $yz,xz$ directions: $\s_z \ct \s_i-\s_i \ct \s_z, \quad i=x,y$~;
	\end{itemize} 
	Hence the deformations that break integrability take the form
	\begin{equation} \label{int.break}
		\begin{split}
			\h_{\text{BInt}}^{(1)}&= \; \delta_{x,+}^{(1)}\;\left(\unit\ct \s_x +\s_x \ct \unit\right)+\delta_{y,+}^{(1)}\;\left(\unit\ct \s_y +\s_y \ct \unit\right)\\
			&+h_{xz,-}^{(1)}\;\left(\s_x \ct \s_z - \s_z \ct \s_x\right)+h_{yz,-}^{(1)}\;\left(\s_y \ct \s_z - \s_z \ct \s_y\right).
		\end{split}
	\end{equation}
	Adding any of these terms to the integrable Hamiltonian will spoil integrability. As expected, we see to first order that deformations either preserve integrability or break it. When we go to the next order we see more subtle distinctions appear in the ways integrability can be broken.
	
	\paragraph{Second order}
	We consider the second order deformation of the integrable case \eqref{def.1st}. We find that there are two possible classes of deformations that preserve integrability. All the restrictions upon $\g^{(2)}$ are explicitly given on the appendix \ref{app.g}.
	The first is given by the following restriction upon $\h^{(1)}$
	\begin{equation}\label{int.o2.1}
		\Delta_{xy}^{(1)}=h_{xy,+}^{(1)}=0
	\end{equation}
	together with the restrictions for $\h^{(2)}$
	\begin{subequations}\label{int.o2.21}
		\begin{align}
			&	h_{xz,-}^{(2)}=-2 \text{coth}\left(\frac{\eta }{2}\right)\; h_{xy,-}^{(1)}h_{yz,+}^{(1)}, \quad h_{yz,-}^{(2)}=2 \text{coth}\left(\frac{\eta }{2}\right)\; h_{xy,-}^{(1)}h_{xz,+}^{(1)},\\[0.25cm]
			&	\delta_{z,+}^{(2)}=2 \text{coth}\left(\frac{\eta }{2}\right)\; h_{xz,+}^{(1)}\delta_{z,+}^{(1)},\quad \delta_{y,+}^{(2)}=2 \text{coth}\left(\frac{\eta }{2}\right)\; h_{yz,+}^{(1)}\delta_{z,+}^{(1)}.
		\end{align}
	\end{subequations}
	We see that imposing integrability of the model imposes extra restrictions on the first order deformation $\h^{(1)}$ given by \eqref{int.o2.1}.
	
	At the end, the first model  that has higher order integrability is given by a deformation at first order and contains 4 free parameters
	\begin{align} \label{def.2nd.a}
		\h_{\text{ModA}}^{(1)}=& \; \delta_{z,+}^{(1)}\;\left(\unit\ct \s_z +\s_z \ct \unit\right)+h_{xz,+}^{(1)}\;\left(\s_x \ct \s_z + \s_z \ct \s_x\right)\\
		+&h_{yz,+}^{(1)}\;\left(\s_y \ct \s_z + \s_z \ct \s_y\right)+h_{xy,-}^{(1)}\;\left(\s_x \ct \s_y - \s_y \ct \s_x\right).
	\end{align}
	It is important to note that this model can be extended to a model that is integrable at second order provided $\h^{(2)}$ satisfies \eqref{int.o2.21}. In fact, it is not hard to show that the second order of Model A can be found by some redefinitions and a basis transformation. One can take the integrable model composed by XXZ, an addition of the spin in the $z$ direction and a DMI term
	\begin{align} \label{mod.1}
		\h_{\text{Int}}=& \; \h_{XXZ}+\eps \delta_{z,+}^{(1)}\;\left(\unit\ct \s_z +\s_z \ct \unit\right)+\eps h_{xy,-}^{(1)}\;\left(\s_x \ct \s_y - \s_y \ct \s_x\right).
	\end{align}
	together with a transformation given by
	\begin{align} \label{gauge.1}
		U=e^{i\eps \text{tanh}\left(\frac{\eta }{2}\right)\left( h_{yz,+}\left(\unit\ct \s_x +\s_x \ct \unit\right))  - h_{xz,+} \left(\unit\ct \s_y +\s_y \ct \unit\right) +\mathcal{O}(\eps^2) \right)}
	\end{align}
	to rewrite
	\begin{align} 
		\h_{\text{ModA}} \left(\eps\right) = U  \left(\eps\right) \h_{\text{Int}}\;{U \left(\eps\right)}^{-1}
	\end{align}
	so Model A can be extended to be integrable at all orders solving recursively for the coefficients in the transformation.
\ysla{The R-matrix for this model can be obtained by applying the basis transformation \eqref{gauge.1} to the R-matrix of the model \eqref{mod.1} given in \cite{beisert_integrable_2013}. }	
	The second model gives rise to a different scenario. We find the following extra restrictions on the first order parameters
	\begin{align}\label{int.o2.2}
		\delta_{z,+}^{(1)}=h_{xy,-}^{(1)}=0
	\end{align}
	and the second order then repeats the pattern given in  \eqref{int.o1.1}
	\begin{equation}\label{int.o2.22}
		h_{xz,-}^{(2)}=h_{yz,-}^{(2)}=0, \qquad \delta_{x,+}^{(2)}=\delta_{y,+}^{(2)}=0.
	\end{equation}
	We find that the second integrable deformation is hence given by
	\begin{equation} \label{def.2nd.b}
		\h_{\text{ModB}}^{(1)}=  \sum_{A,B = x,y,z} h_{AB,+}^{(1)}\left(\s_A \ct \s_B + \s_B \ct \s_A\right).
	\end{equation}
	In particular due to the recurring structures we find that the full integrable model of which \eqref{def.2nd.b} is the first order of is simply given by
	\begin{align} 
		\h_{\text{ModB}} = \h_{XXZ} + \sum_{A,B = x,y,z} h_{AB}(\epsilon)\left(\s_A \ct \s_B + \s_B \ct \s_A\right),
	\end{align}
	where $h_{AB}(\epsilon)$ is an arbitrary power series in $\eps$. It can be readily checked that this model satisfies the integrability constraint coming from the boost operator formalism. \ysla{This would represent a generic rotated XYZ model, so its R-matrix is expected to be related to the usual eight-vertex R-matrix by a rotation.Furthermore, as the full classification of regular $4 \times 4$ $R$-matrices has been achieved \cite{Corcoran:2023zax}, the $R$-matrices for both Model A and Model B are contained within this classification.}
	
	\paragraph{Quasi-integrable models}
	In the same manner, we can define quasi-integrable models as those that respect \eqref{int.o1.1} but satisfy neither the conditions \eqref{int.o2.1} nor \eqref{int.o2.2}. This gives a complete classification of the quasi-integrable models that are only integrable to first order in $\eps$. In order to work with a concrete model in what follows, we consider the parameterisation
	\begin{align} \label{quasi.i}
		H_{\text{QInt}}&=H_{\text{XXZ}}+\epsilon \alpha \sum_{i=1}^L (\sigma_{x,i} \sigma_{x,i+1}-\sigma_{y,i} \sigma_{y,i+1})\\
		&+\epsilon \beta \sum_{i=1}^L \left(\sigma_{z, i}+\sigma_{z,i+1}\right)+\epsilon \gamma  \sum_{i=1}^L (\sigma_{x,i}\sigma_{y,i+1}-\sigma_{y, i}\sigma_{x, i+1}). \nonumber
	\end{align}
	This is a deformation of the XXZ spin chain including a total spin operator, an XYZ term and a DMI term.

	\section{Perturbative $R$-matrices}
	\label{sec:Rmat}
	Let us now consider what happens on the level of the $R$-matrix and see whether the deformations come from a solution of the Yang--Baxter equation.
	
	\paragraph{Sutherland equation}
	When the Hamiltonian is known, the $R$-matrix can be obtained from the Sutherland equation
	\begin{equation} \label{sutherland}
		\left[R_{13}(u,v) R_{23}(u,v), \h_{12}(u)\right]=\pa_u R_{13}(u,v) R_{23}(u,v)-  R_{13}(u,v)\pa_u R_{23}(u,v).
	\end{equation}
	Given that we expand our Hamiltonian in our deformation parameter $\eps$, we do the same for the $R$-matrix and write
	\begin{equation}
		R_{ij} = R_{ij}^{(0)} +\eps R_{ij}^{(1)} + \mathcal{O}(\eps^2)
	\end{equation}
	and solve the Sutherland equation order by order. In our case we perturb the XXZ spin chain whose $R$-matrix is well-known and given by
	\begin{align} \label{r0.st}
		R^{(0)}(u)=
		\csch(\eta)
		\left(
		\begin{array}{cccc}
			\sinh(u+\eta ) & 0 & 0 & 0 \\
			0 & \sinh(u) & \sinh(\eta) & 0 \\
			0 & \sinh(\eta)  & \sinh(u) & 0 \\
			0 & 0 & 0 & \sinh(u+\eta) \\
		\end{array}
		\right).
	\end{align}
	It is easily checked that the logarithmic derivative reproduces the Hamiltonian \eqref{HXXZ}.
	
	For the integrable deformations we can obviously find a solution to any order in $\eps$. Similarly, for the integrability breaking solutions, we will not be able to find an $R$-matrix. So, finally, let us look at the quasi-integrable model \eqref{quasi.i}. For simplicity, let us set $\beta=0$ in which case $\g=0$, \ysla{following the relations given by  \eqref{g.rel}} .This means that the $R$-matrix will be of difference form. We thus write
	\begin{equation}
		R(u) = R_{XXZ}\left(u\right) +\eps R^{(1)}(u) + \mathcal{O}(\eps^2)
	\end{equation}
	for a  $4 \times 4$  correction to be determined. Substituting into the Sutherland equation and solving the resulting PDEs for $R_{ij}^{(1)}\left(u\right)$ at first order in $\eps$, we find the following solution
	\begin{align}
		R^{(1)}(u)=  \sinh(u-v)  \left(
		\begin{array}{cccc}
			0& 0 & 0 & 2\alpha  \frac{\sinh (u-v+\eta) }{\sinh(\eta)}\\
			0 & -2i \gamma  & 0 & 0 \\
			0 & 0 & 2i \gamma  & 0 \\
			2 \alpha \frac{ \sinh (u-v+\eta)}{\sinh(\eta)} & 0 & 0 & 0\\
		\end{array}
		\right).      	
	\end{align}
	We directly verified that this R-matrix satisfies Yang-Baxter equation to first order in $\eps$. One could propose to continue the expansion to higher order, i.e, we write
	\begin{equation}
		R = R^{(0)} +\eps R^{(1)}+\eps^2 R^{(2)}
	\end{equation}
	and solve to second order. As expected, solving the second order Sutherland equation we  obtain equations that do not involve the functions $R^{(2)}_{ij}$ but only depend on the entries of the Hamiltonian.
	In this way we recover the equations \eqref{int.o2.1} or \eqref{int.o2.2} that lead us back to the integrable cases.
	
	\paragraph{Lax operator}
	When $\beta\neq0$ and we have a non-trivial $\g$, this means that the corresponding $R$-matrix is of non-difference form and the coefficients of the Hamiltonian generically depend on the spectral parameter. In this case it is more convenient to follow \cite{deLeeuw:2022ath} and construct the Lax matrix instead.
	
	For a given deformation, we know that in order to be regular and produce the correct Hamiltonian the Lax operator has to take the following form
	\begin{align}
		\mathcal{L}_{QInt}(u) = \mathcal{L}_{XXZ}(u) + \eps \mathcal{P} \;\left(\sinh(u) \h^{(1)} - \Lambda(u)\right),
	\end{align}
	where $\mathcal{P}$ is the permutation operator and $\Lambda(u)$ is the correction term that we need to determine. It needs to satisfy $\Lambda(0) = \Lambda'(0)  = 0$. From this Lax operator we can then compute the corresponding transfer matrix and demand that it commute with the Hamiltonian. Taking into account the boundary conditions, we are then lead to the solution
	\begin{align}
		\Lambda = \sinh(u) \left(
		\begin{array}{cccc}
			0 & 0 & 0 & 2\alpha  (1-\frac{\sinh (u+ \eta )}{\sinh(\eta)}) \\
			0 & 2\beta  \frac{\sinh (2 u)}{\sinh(\eta)} & 0 & 0 \\
			0 & 0 & 0 & 0 \\
			2\alpha  (1-\frac{\sinh (u+ \eta )}{\sinh(\eta)}) & 0 & 0 & 4 \beta  \frac{\sinh (u)}{\sinh(\eta)} \cosh (\eta +u) \\
		\end{array}
		\right).
	\end{align}
	This Lax operator satisfies the RLL relations up to order $\eps$ with $R$-matrix
	\begin{align}
		R  = R_{XXZ} + \eps \sinh(u-v) R^{(1)},
	\end{align}
	with
	\begin{align}
		R^{(1)}=
		\left(
		\begin{array}{cccc}
			-2\beta  \frac{\sinh (2 v-\eta )}{\sinh(\eta)} & 0 & 0 & 2\alpha  \frac{\sinh (u-v+\eta) }{\sinh(\eta)}\\
			0 & -2\beta  \frac{\sinh (2 u)+\sinh (2 v)}{\sinh(\eta)}-2i \gamma  & 0 & 0 \\
			0 & 0 & 2i \gamma  & 0 \\
			2\alpha \frac{ \sinh (u-v+\eta)}{\sinh(\eta)} & 0 & 0 & -2\beta  \frac{ \sinh (2 u+\eta) }{\sinh(\eta)}\\
		\end{array}
		\right).
	\end{align}
	We see that this $R$-matrix is indeed of non-difference form and it can be easily checked that it satisfies the Yang--Baxter equation and braiding unitarity up to first order in $\eps$. It can again be easily checked that this $R$-matrix can not be extended to a solution of the Yang--Baxter equation to second order in $\eps$ unless the parameters $\alpha,\beta,\gamma$ satisfy additional constraints that bring us back to the integrable case.

	\section{Spectral Statistics}
	\label{sec:stats}
	We now turn to a numerical analysis of the deformed models introduced above. We will compute indicators of integrability breaking and the onset of quantum chaos as we increase the deformation parameter $\epsilon$ away from the integrable point. We will take numerical representatives from the quasi-integrable model $H_{\text{QInt}}$, see \eqref{quasi.i}, and for comparison, from a model which exhibits integrability breaking. We do not consider the models for which the integrability condition can be solved to higher order as they can be written as trivial restrictions of integrable models and do not demonstrate any onset of chaos (unlike the long-range deformations studied in \cite{Szasz-Schagrin:2021pqg, McLoughlin:2022jyt}).
	In this section, we will make use of the parameterisation 
	\begin{align}
		H_{\text{XYZ}}= \sum_{i=1}^L \big[J_x ~\sigma_{x,i}\sigma_{x,i+1}+J_y~ \sigma_{y,i}\sigma_{y,i+1}+J_z~ \sigma_{z,i}\sigma_{z,i+1}\big]
	\end{align}
	with $J_x=J_y=1$ for the XXZ model. 
	%
	%\paragraph{Quasi-integrable model}
	%For our quasi-integrable Hamiltonian we take  following Hamiltonian
	%%
	%\begin{align}
	%H_{\text{QInt}}=H_{\text{XXZ}}+\epsilon \alpha \sum_{i=1}^L (\sigma_{x,i} \sigma_{x,i+1}-\sigma_{y,i} \sigma_{y,i+1})+\epsilon \beta \sum_{i=1}^L \sigma_{z, i}+\epsilon \gamma  \sum_{i=1}^L (\sigma_{x,i}\sigma_{y,i+1}-\sigma_{y, i}\sigma_{x, i+1}).
	%\end{align}
	%%
	%This Hamiltonian reduces to the integrable XXZ spin-chain for $\epsilon=0$ but is only integrable to $\mathcal{O}(\epsilon)$ when all components are non-zero. 
	For our prototype of an integrability-breaking Hamiltonian we consider the XYZ Hamiltonian deformed by the DMI interaction term
	\begin{align}
		H_{\text{dXYZ}}=H_{\text{XYZ}}+\epsilon  \sum_{i=1}^L (\sigma_{x,i}\sigma_{y,i+1}-\sigma_{y, i}\sigma_{x, i+1})~.
	\end{align}
	This deformation has the advantage of preserving the same global symmetries as $H_{\text{QInt}}$. 
	
	%
	%We can consider the spectrum of the deformed Hamiltonians listed in section \ref{sec:models} numerically in order to study their statistical behaviour.
	
	\paragraph{Symmetry sectors}
	To analyse the spectral statistics a necessary first step is appropriately decomposing the Hamiltonian into independent symmetry sectors. Ideally, one should remain in the same symmetry sector for all values of the deformation. The XXZ spin-chain is invariant under the standard lattice symmetries, translations and lattice parity, plus $U(1)\rtimes \mathbbm{Z}_2$ unitary symmetries corresponding to rotations about the $z$-axis and $\mathbbm{Z}_2$ spin-flips generated by $F_2=\prod_i \sigma_{x,i}$ which sends $\{\sigma_z, \sigma_y\}\to \{-\sigma_z, -\sigma_y\}$.
	However, the combination of XYZ and total spin-operator deformations, corresponding to the parameters $\alpha$ and $\beta$ in $H_{\text{QInt}}$, \eqref{quasi.i}, break the global $U(1)$ symmetry down to a $\mathbbm{Z}_2$ generated by $F_1=\prod_i \sigma_{z,i}$. The DMI deformation corresponding to $\gamma$ breaks the $F_2$ spin-flip symmetry as well as lattice parity. We must thus focus on the sectors corresponding to these remaining symmetries. 
	
	As we consider periodic chains, translation invariance results in a spectrum which decomposes into sectors of definite lattice momentum. Generally in such analysis it is most convenient to consider the translationally invariant, momentum zero, sector and further decompose into sub-sectors of fixed parity. However, as parity is broken by the $\gamma$-deformation we instead consider sectors of non-zero momentum. In the undeformed XXZ model, parity maps these $p\neq 0$ momentum sectors to $-p$, so that even in the $\epsilon\to 0$ limit there are no additional degeneracies. 
	Further, the XYZ deformation mixes sectors of different total impurity number $M$, where $M=L/2-S_z$, and so we must consider sectors of different magnetisation eigenvalues. However, it does preserve the impurity number mod 2 and so we restrict to sectors of even impurity number, which corresponds to a specific eigenspace of $F_1$. 
	
	While the deformations break much of the symmetry of the undeformed model there remains an anti-unitary symmetry corresponding to a combination of a lattice parity transformation and complex conjugation which leaves the Hamiltonian invariant. This results in the spectral statistics being GOE as we will see below. 
	
	\subsection{Level Spacing}
\begin{figure}
		\vspace{0.3cm}
		\begin{center}
			\includegraphics[height=0.3\linewidth]{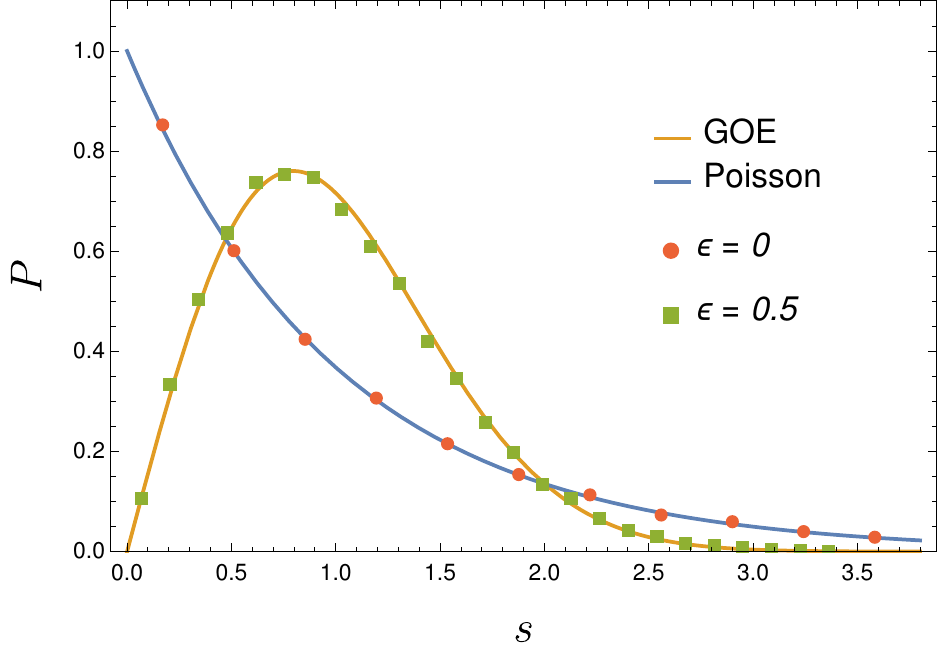}
			\hspace{3mm}
			\includegraphics[height=0.3\linewidth]{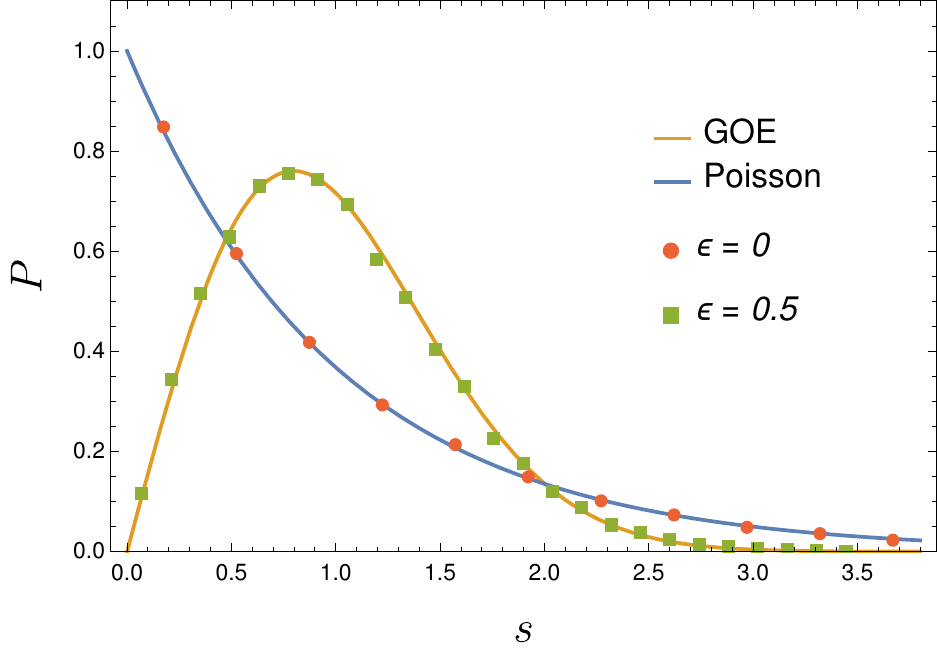}
		\end{center}
		\caption{Level-spacing distributions for integrable, $\epsilon=0.0$, and chaotic, $\epsilon=0.5$, regimes in $L=21$ chain in momentum-one sector with even magnetisation which has dimension 49,929 for (Left) $H_{\text{QInt}}$ with  $J_z=0.68$, $\alpha=0.82$, $\beta=1.43$, $\gamma=1$.   (Right) $H_{\text{dXYZ}}$ with $J_x=1.63, J_y= 0.73, J_z =1.18$.  }	\label{fig:quasi_spectral}
	\end{figure}

	To analyse the effects of the deformations we choose representative values for the spin-chain parameters and then numerically diagonalise the Hamiltonian. For comparison with the predictions of random matrix theory we must focus on the fluctuations about the mean eigenvalue density and so we must remove the overall smooth dependence by ``unfolding" the spectrum such that the unfolded values have a uniform mean density. Furthermore, as is typical for spin-chain spectra, the eigenvalues near the edges of the spectrum are non-generic, for example they do not show chaotic behaviour, and so we remove a portion from the spectrum. We provide some analysis of the effects of specfic numerical choices in App. \ref{app:clipping}. This reflects the fact that the transition to chaos is not uniform across the spectrum and different states transition at varying rates. From the unfolded eigenvalues $\epsilon_k$ we compute the level spacings $s_k=\epsilon_{k+1}-\epsilon_k$ and, by grouping them into bins, approximate their probability distribution. For an integrable model we expect the distribution of spacings to be Poisson
	\begin{align}
		P_P(s)=e^{-s}
	\end{align}
	while for a chaotic system we expect the distribution to closely follow Wigner's surmise. For systems where the Hamiltonian can be chosen real and symmetric this will be given by the GOE Wigner-Dyson distribution
	\begin{align}
		P_{GOE}(s)=\tfrac{1}{2} \pi se^{-\frac{\pi}{4} s^2}~.
	\end{align}
	We compare these predictions with the results of our numerical calculations for $H_{\text{dXYZ}}$ and $H_{\text{QInt}}$ in Fig. (\ref{fig:quasi_spectral}). 
	As can be seen, the statistics for both models are Poisson in the $\epsilon=0$ case and Wigner-Dyson for sufficiently large deformation, $\epsilon=0.5$. At these extremes of the coupling range the models behave essentially identically from the perspective of the level spacing distribution.
	
\begin{figure}
		\vspace{0.3cm}
		\begin{center}
			\includegraphics[height=0.3\linewidth]{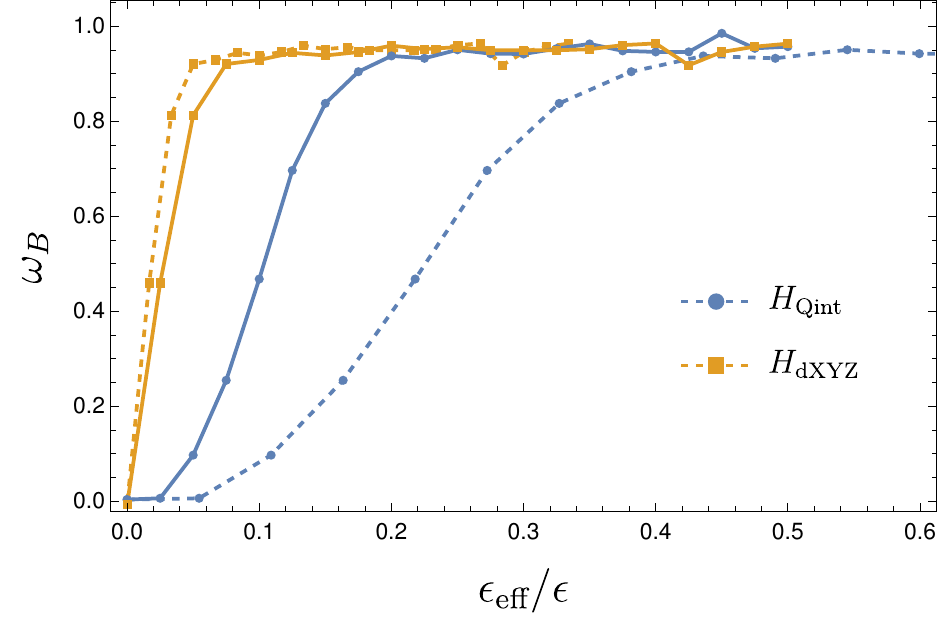}
			\hspace{3mm}
			\includegraphics[height=0.3\linewidth]{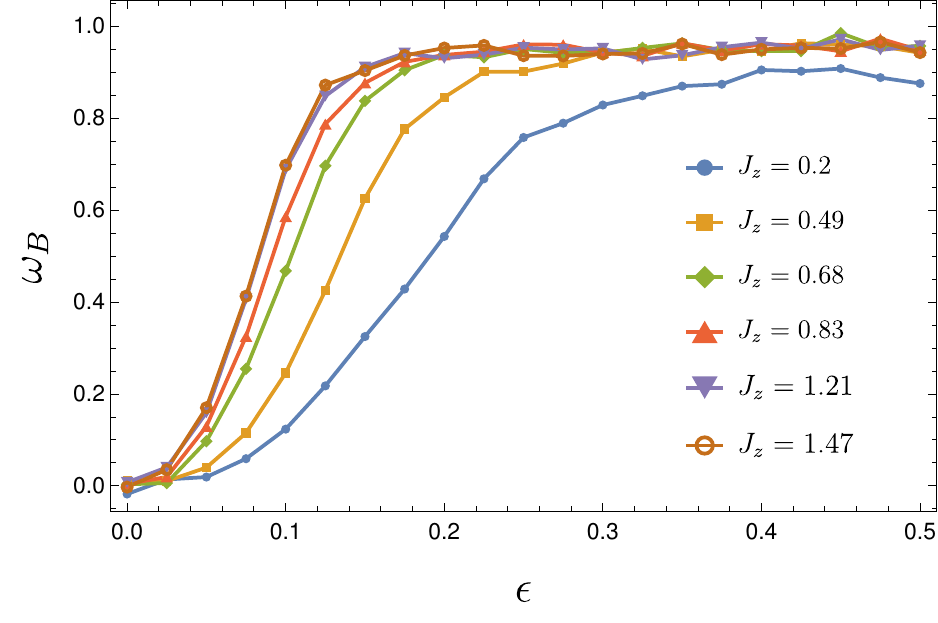}
		\end{center}
		\caption{Brody parameter $\omega_B$ as a function of deformation strength  for $L=21$ in the momentum-one, even-magnetisation sector (dim. 49,929) for (Left) $H_{\text{QInt}}$ with  $J_z=0.68$, $\alpha=0.82$, $\beta=1.43$, $\gamma=1$ and $H_{\text{dXYZ}}$ with $J_x= 1.63, J_y= 0.73, J_z =1.18$. Dashed lines are for a rescaled, effective coupling. (Right) $H_{\text{QInt}}$ for  $\sigma_{xy}=1$, $\alpha=0.82$, $\beta=1.43$, $\gamma=1$ and different values of $J_z$.  }	\label{fig:trans_L21}
	\end{figure}

	However,  the changeover from one regime to the other is quite different for the two models. To study how this takes place we compute the spectrum for a range of values $\epsilon\in [0, 0.5]$. We then fit the resulting level spacings to the phenomenologically inspired Brody distribution
	\begin{align}
		P_{\text{Brody}}(s)=(\omega_B +1)\Gamma \left(\frac{\omega_B+2}{\omega_B +1}\right)^{\omega_B +1}  s^{\omega_B }  e^{- \Gamma \left(\frac{\omega_B+2}{\omega_B +1}\right)^{\omega_B
				+1} s^{\omega_B +1}}
	\end{align}
	which is Poisson for $\omega_B=0$ and GOE WD for $\omega_B=1$. The resulting values of $\omega_B$ as a function of $\epsilon$ are shown in Fig. (\ref{fig:trans_L21}). As can be seen, the behaviour for the two models is quite distinct with the onset of chaos occurring at larger values of $\epsilon$ for the quasi-integrable model $H_{\text{QInt}}$ compared to $H_{\text{dXYZ}}$.
	The effect is in fact even more pronounced when the sizes of the relative deformations are taken into account.
	Quantitative results can be found in App. \ref{app:norms} but qualitatively, as the deformation in $H_{\text{QInt}}$ involves more terms than that of $H_{\text{dXYZ}}$ it has a larger norm. Thus one might naively expect a stronger indicator of chaos in $H_{\text{QInt}}$ for a given $\epsilon$ and the fact that the opposite occurs is presumably due to the quasi-integrability of the model. There is also a difference in the shape of the transitions. For $H_{\text{dXYZ}}$, the parameter  $\omega_B$ shows a immediate linear increase which persists until close to the plateau. This initial linear increase seems to be common in related chaotic spin-chains such as those with generic next-to-nearest neighbour (NNN) interactions. For $H_{\text{QInt}}$ there is initially only a slow increase, and possibly even a region where $\omega_B$ remains at zero, before a more rapid increase occurs until close to the plateau. 
	
\begin{figure}
		\vspace{0.3cm}
		\begin{center}
			\includegraphics[height=0.3\linewidth]{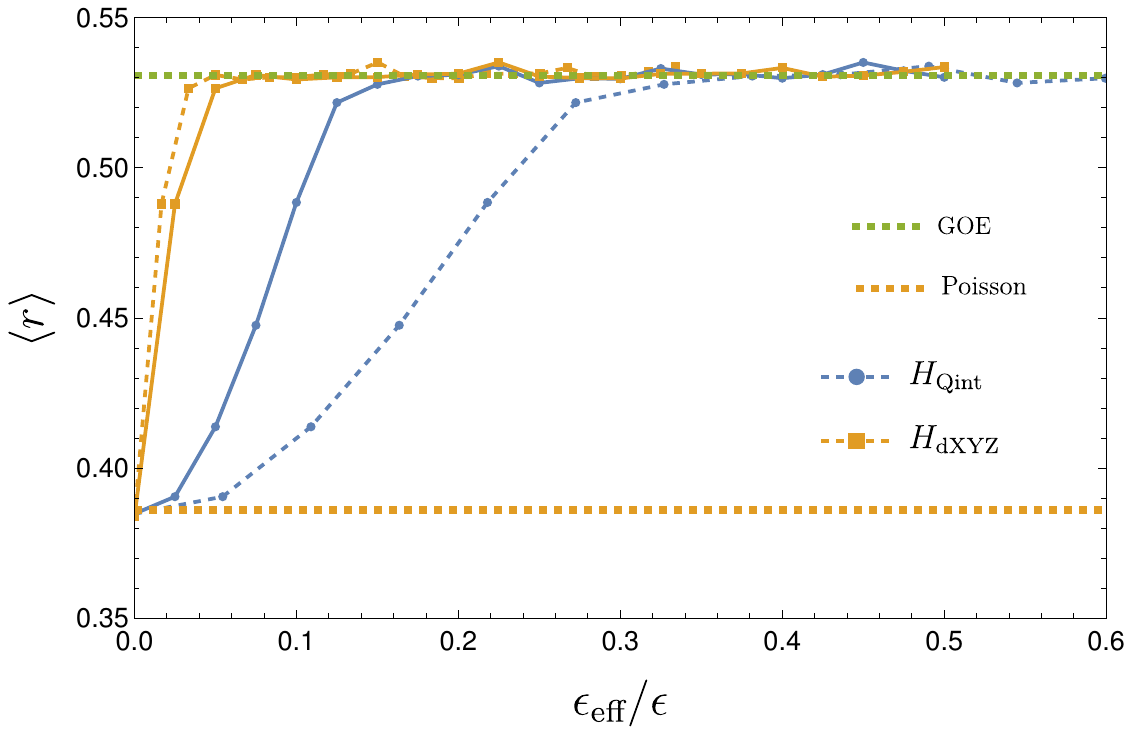}
			\hspace{3mm}
			\includegraphics[height=0.3\linewidth]{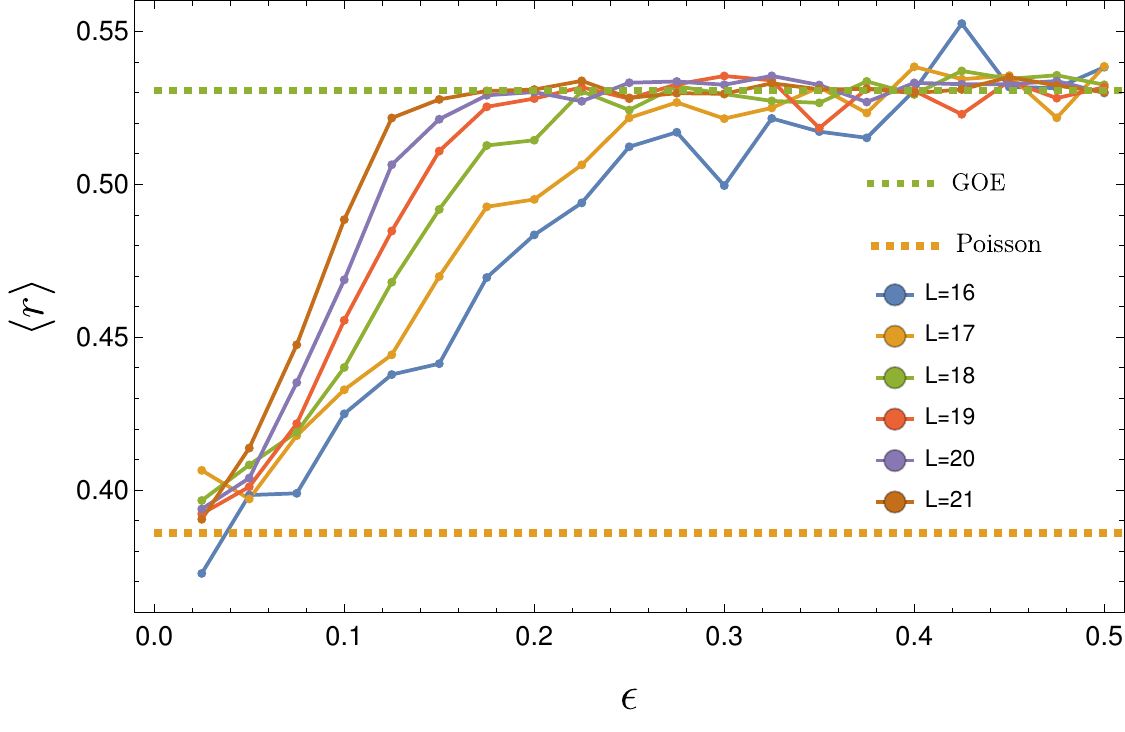}
		\end{center}
		\caption{(Left) $r$-ratio as a function of deformation strength  for $L=21$ in the momentum-one, even-magnetisation sector for $H_{\text{dXYZ}}$ with $J_x= 1.63, J_y= 0.73, J_z =1.18$ and $H_{\text{QInt}}$ with  $J_z=0.68$, $\alpha=0.82$, $\beta=1.43$, $\gamma=1$. Dashed lines are for a rescaled, effective coupling. (Right) $r$-ratio for $H_{\text{QInt}}$ with $J_z=0.68$, $\alpha=0.82$, $\beta=1.43$, $\gamma=1$ for $L\in [16,21]$.  }	\label{fig:rtrans}
	\end{figure}

Similar behaviour appears in other spectral measures, including quantities that do not require unfolding, which provides a check that our conclusions are robust to the details of the numerical procedure. For example, the ratio of consecutive level spacings $\tilde{r}_i=S_i/S_{i-1}$ can be defined using the raw energy eigenvalues $S_i=E_{i+1}-E_i$ as the overall density cancels in the ratio. We can then introduce the quantity \cite{Oganesyan:2007wpd}, $r_i=\min \left(\tilde{r}_i, \frac{1}{\tilde{r}_i}\right)$ whose mean value is known to interpolate between 0.531 for the GOE distribution and 0.387 for Poisson distributed eigenvalues. In Fig. (\ref{fig:rtrans}) left panel, we see that, just as the Brody parameter interpolates from $0$ to $1$ with increasing deformation parameter, $\langle r\rangle$ interpolates between the Poisson and GOE values. Moreover, it shows very similar behaviour across models with a slower increase for $H_{\text{QInt}}$ compared to $H_{\text{dXYZ}}$.
	
	It is also of interest to consider the dependence of the changeover on the XXZ anistropy parameter. For a relatively wide range of values, $J_z\gtrsim 0.5$ the dependence is quite weak as can be seen in Fig. (\ref{fig:trans_L21}). In some regards this is slightly surprising. One might expect that as the norm of $H_{\text{XXZ}}$ decreases with decreasing $J_z$, see App. \ref{app:norms} for numerical results, and so the relative size of the deformation is larger, one might see a faster transition but this is not reflected in the numerics which instead show a slight slowing. Additionally, there does not seem to be a strong dependence on whether the undeformed model is in the gapped $|J_z|>1$ or gapless phase $|J_z|<1$. The dependence becomes more pronounced at smaller values, e.g. $J_z=0.2$ where $H_{\text{QInt}}$ does not become strongly chaotic for any value of the deformation. This is possibly due to being relatively close to the free fermion, $J_z=0$, point. For $J_z\lesssim 0.5$ the anisotropy term is of a comparable size to the other deformations when $\epsilon\simeq 0.5$ and so we should view this as a deformation of the XY model. 
	
	\subsection{Volume dependence of critical coupling}
	\begin{figure}
		\vspace{0.3cm}
		\begin{center}
			\includegraphics[height=0.3\linewidth]{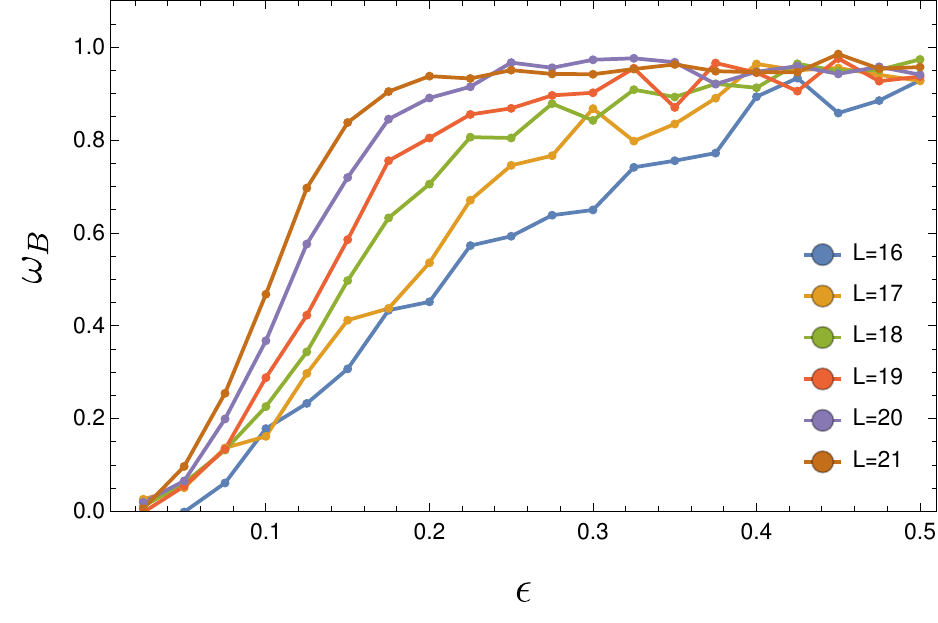}
						\hspace{3mm}
			\includegraphics[height=0.3\linewidth]{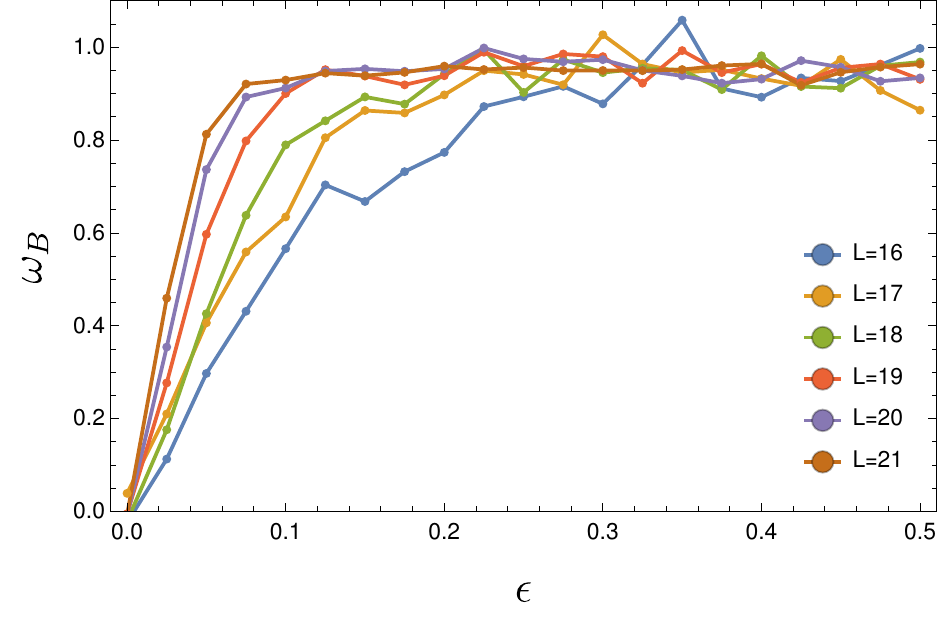}
		\end{center}
		\caption{Brody parameter $\omega_B$ as a function of deformation strength $\epsilon$ for $L\in [16,21]$ in the momentum-one, even magnetisation sector. (Left)  $H_{\text{QInt}}$ with $J_z=0.68$, $\alpha=0.82$, $\beta=1.43$, $\gamma=1$. (Right) $H_{\text{dXYZ}}$ with $J_x= 1.63, J_y= 0.73, J_z =1.18$}	\label{fig:trans_scaling}
	\end{figure}
	The dependence of the transition on the length of the spin-chain $L$ provides a finer grained view of this process. In Fig. \ref{fig:trans_scaling} we plot the transition curves for the dXYZ and QInt models as a function of $\epsilon$ for different values of the length.  For the dXYZ model there is a clear pattern where the initial slope increases with increasing $L$ and so the transition between regimes occurs for progressively smaller values of the deformation parameter. For the QInt model there is a similar steepening of the curve but the initial slow take-off remains\footnote{It should be noted that for the QInt model in Fig. \ref{fig:trans_scaling} we do not include the $\epsilon=0$ point. This is a point of enhanced symmetry, where the $U(1)$ rotational symmetry is restored, such that the spectrum further decomposes into magnetisation sub-sectors which are statistically independent. As can be seen the distribution is still very close to Poisson for small, $\epsilon\simeq 0.025$, values of the deformation parameter.}. A very similar dependence on the spin-chain length is also seen in the mean ratio of consecutive spacings, see Fig. (\ref{fig:rtrans}) right panel.
	
\begin{figure}
		\vspace{0.3cm}
		\begin{center}
			\includegraphics[height=0.3\linewidth]{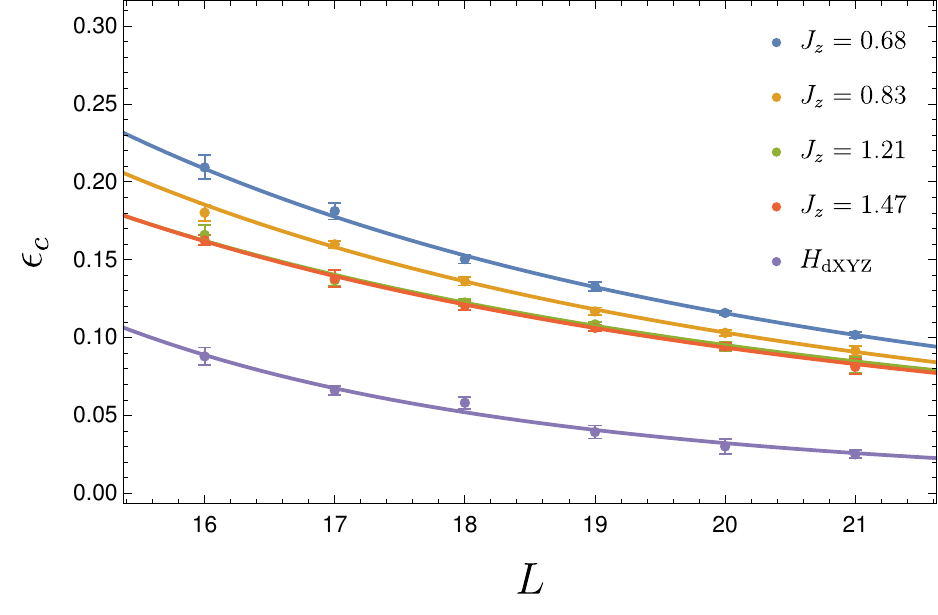}
			\hspace{3mm}
			\includegraphics[height=0.3\linewidth]{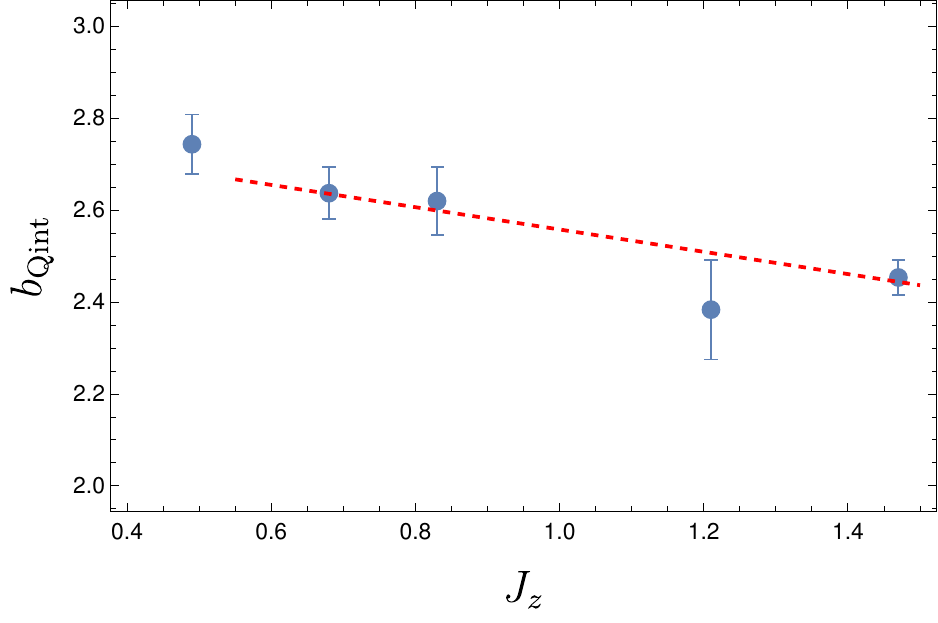}
		\end{center}
		\caption{(Left) Scaling of critical coupling $\epsilon_c$ as a function of $L$ in the momentum-one, even-magnetisation sector of the  QInt model $\alpha=0.82$, $\beta=1.43$, $\gamma=1$ for different values of $J_z$ and the dXYZ model, with $J_x= 1.63, J_y= 0.73, J_z =1.18$. Curves are best-fit power-law scaling. (Right) Values of power-law scaling exponent $b$ for QInt model for values of $J_z$. Dashed line is a linear fit to the last four points.}	\label{fig:crit_scaling}
	\end{figure}

	Given this scaling, one can define a critical coupling $\epsilon_c$ where the distribution is mid-way between integrable and chaotic i.e. $\omega_B=1/2$. \footnote{Technically, we extract the $\epsilon_c$ by fitting the transition data to a degree-six polynomial and solving for $\omega_B=1/2$. The errors are estimated using a ``leave-one-out" approach of fitting to restricted data sets and computing the r.m.s. difference.}
	This critical coupling decreases with increasing $L$ for both models as can be seen in Fig. (\ref{fig:crit_scaling}). 
	The exact dependence of $\epsilon_c$ on the size of the system is not currently well-understood. In \cite{modak2014universal, modak2014finite}, the authors proposed that for gapless systems $\epsilon_c$ scales as $L^{-3}$ while for systems with a gap they noted a faster scaling on the system size.  Power-law results with smaller exponents, $L^{-2}$ and $L^{-1}$, were found in gapless, ``weakly"-chaotic systems \cite{Szasz-Schagrin:2021pqg, McLoughlin:2022jyt}. The authors of \cite{bulchandani2021onset} proposed similar scalings, $1/(L^{n}\ln L)$ for \textit{transitions}, generated by perturbations which correspond to short-range interactions in Fock space while for \textit{crossovers}, generated by perturbations which are long-range in Fock space, they demonstrated exponential scaling of the critical coupling.  
	
	For the QInt model with, for example, $J_z=0.68$ if we fit to a power-law scaling
	\begin{align}
		\epsilon_c=\frac{a}{L^b}
	\end{align}
	we find that 
	\begin{align}
		a_{\text{QInt}}=(3.1\pm 0.5)\times 10^2~,~~~b_{\text{QInt}}=2.64 \pm 0.06
	\end{align}
	which clearly lies between the generic result of $b=3$ and the weakly-chaotic result of $b=2$. Fitting to a functional form $\frac{a}{L^b\ln L}$ results in similar numerical parameters ($a=(3.4\pm 0.6)\times 10^2$, $b=2.30\pm 0.06$).
	Changing the value of $J_z$ results in relatively small changes to this parameter, with the values centred around $b\simeq 2.5$, though there is a general trend of decreasing $b$ with increasing $J_z$, see Fig. (\ref{fig:crit_scaling}). For sufficiently small values of $J_z$ the system never obtains a parameter $\omega_B=1/2$ at shorter lengths and we cannot cleanly analyze the transition. For this reason we restrict to values $J_z\gtrsim 0.5$. We do not see a clear change in behaviour as we cross into the gapped regime $|J_z|>1$. 
	While the fit to a power-law is clearly quite reasonable, given the limited range of numerically accessible lengths  it is by no means uniquely determined.  
	Fitting to the exponential functional form
	\begin{align}
		\epsilon_c= ce^{-d L}
	\end{align}
	we find, again with $J_z=0.68$, that $c_{\text{QInt}}=1.9\pm 0.2$ and $d_{\text{QInt}}=0.140\pm 0.005$. Similar values around $d_{\text{QInt}}\sim 0.14$ are found for differing values of $J_z$. Thus, in this case it is hard to determine whether, in the terminology of \cite{bulchandani2021onset}, we are seeing a transition or a crossover.  However, while we have not been able to fix the form of the scaling, it does seem quite clear that the quasi-integrable model provides an example of an intermediate type theory with scaling between the strong-breaking and weakly-breaking results. 
	
	For comparison we can consider the dXYZ model. If we fit to a power-law scaling
	we find that $a_{\text{dXYZ}}=(2.5\pm 2.4)\times 10^4$, $b_{\text{dXYZ}}=4.5\pm 0.2$ which is faster than the generic result of $b=3$ \cite{modak2014universal, modak2014finite}. However, this fit is not particularly good with the error in $a$ being $\mathcal{O}(a)$. Fitting to the functional form 
	$\epsilon_c= ce^{-d L}$
	we find 
	\begin{align}
		c_{\text{dXYZ}}=5 \pm 1~~~ \text{and}~~~ d_{\text{dXYZ}}=0.25\pm 0.02.
	\end{align}
	which while still not a particularly good fit is significantly better. This suggests that this is better described as a crossover with an exponential scaling of the critical coupling. 

	\subsection{Eigenvector entropy}
	\label{sec:entropy}
	
	For a truly random matrix ensemble eigenvectors of different Hamiltonians should be uncorrelated and so entirely delocalised when expressed in terms of each other. However, in physical systems this de-localisation is generally incomplete and only bounded from above by the RMT result. We can quantify this by computing the information entropy for eigenvectors
	\begin{align}
		S^{(k)}=-\sum_{n=1}^D |c^{(k)}_n|^2 \ln |c^{(k)}_n|^2
	\end{align}
	where each eigenvector can be decomposed $\ket{E_k}=\sum_{n=1}^D c_n \ket{n}$ and $\{\ket{n}\}$ is an orthonormal basis. The information entropy acts as an additional measure of ergodicity in the transition region, but also provides information regarding the structure of the system as a function of energy. In this later sense it acts as a step towards understanding the validity of ETH in the model. 
	
\begin{figure}
		\vspace{0.3cm}
		\begin{center}
			\includegraphics[height=0.3\linewidth]{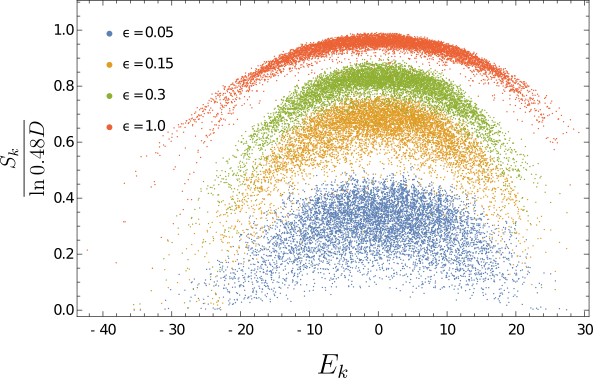}
			\hspace{3mm}
			\includegraphics[height=0.3\linewidth]{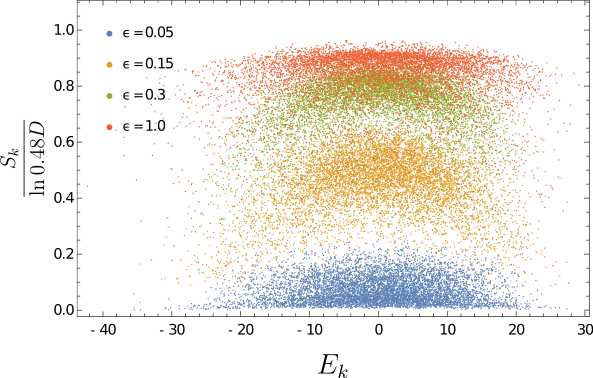}\\
			\includegraphics[height=0.3\linewidth]{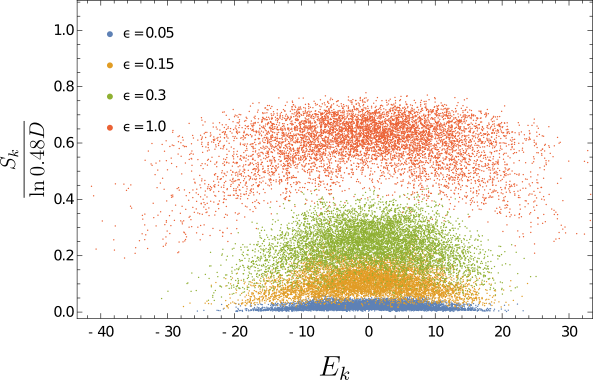}
			\hspace{3mm}
			\includegraphics[height=0.3\linewidth]{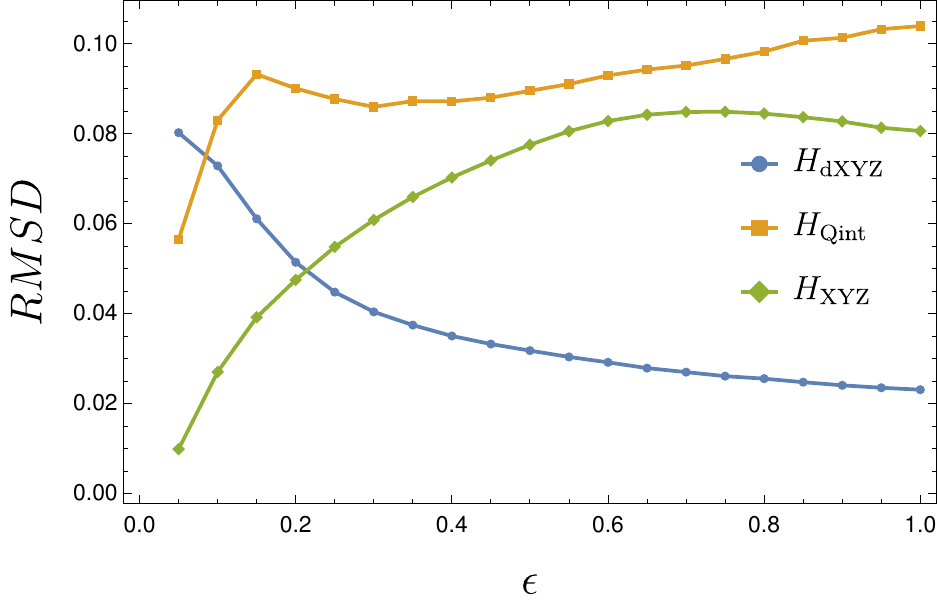}
		\end{center}
		\caption{ Information entropy for states in the $L=18$ in the momentum-one, even-magnetisation sector (dim$=7252$) of the (Top left) deformed XYZ model with $\Delta_{xy}=0.45$, $\sigma_{xy}=J_z=1.18$, and coupling $\epsilon$ (Top right) QInt model with $\sigma_{xy}=1$, $J_z=0.68$, $\gamma=1$, $\alpha=0.82$, $\beta=1.4$ (Bottom left) XYZ model with $\Delta_{xy}=0.45$, $\sigma_{xy}=0.59+1.18 \epsilon$, $J_z=1.18$. (Bottom right) R.M.S. of differences to a fitted quadratic curve for the three models. }	\label{fig:entropy_fits}
	\end{figure}

	We take the eigenstates for the undeformed model as the reference basis and then compute the entropy of the eigenstates as we increase the deformation parameter. For a deformation which breaks integrability, as in the dXYZ model, we can see in Fig. \ref{fig:entropy_fits} (Top left) that as the deformation grows the states rapidly delocalise and the entropy generally increases until in the middle of the spectrum it is quite close, $\simeq 96\%$, to the GOE prediction of $S_{\text{GOE}}=\ln(0.48 D)$. Towards the edges of the spectrum it however remains significantly lower. For comparison we can see, Fig. \ref{fig:entropy_fits} (Bottom left), that for an integrable deformation, we consider the XYZ model and vary $\sigma_{xy}$, the entropy has a slower increase and the maximum value is far from the RMT prediction, $\simeq 61\%$. Between these extremes, the quasi-integrable QInt model shows, Fig. \ref{fig:entropy_fits} (Top right), increasing entropy, with a maximum value that is comparable with the dXYZ models,  $\simeq 97\%$ of RMT. However, it increases slightly slower with $\epsilon$ than the integrability breaking deformation. 

	Perhaps more significantly, for the integrability breaking deformation, the entropy becomes a relatively smooth function of the energy as one would expect in a thermalised system. This is not true of the integrable case or the quasi-integrable deformation. We can characterise this by fitting the entropy to a smooth curve (we choose a quadratic function) and computing the r.m.s. of the differences (RMSD) between the estimated curve and the data points. This is shown in Fig. \ref{fig:entropy_fits} (Bottom right) where the RMSD for the dXYZ model decreases with increasing deformation while for the XYZ and QInt models the variance is relatively high and shows little sign of decreasing.

	\section{Conclusions and Discussion}
	
	In this paper we studied the nearest-neighbour deformations of the XXZ spin chain by using the boost operator formalism. After expanding in the deformation parameter, we identified four possible types of deformations. 
	There are deformations that simply break or preserve integrability, but there are also cases in which the integrability only holds to a certain order. We saw that there are still two different possibilities. Either
	the deformation can be extended to a proper integrable model if higher orders in the deformation parameter are taken into account or this is not possible.
	
	The numerical studies of the spectral statistics in section \ref{sec:stats} confirm this picture and indicate that it is relevant to the physical properties of the models such as the onset of chaos. The models with broken integrability, $H_{\text{dXYZ}}$, show a relatively sharp onset of chaos with increasing coupling $\epsilon$ parameter and a scaling of the critical coupling that is best described as exponential $\epsilon_c \sim e^{-d L}$. This is consistent with the scaling proposed in \cite{modak2014finite, modak2014universal} due to the gapped nature of the XYZ model and with the crossover type behaviour of \cite{bulchandani2021onset} suggesting a non-local process in Fock space. The energy eigenstates show an increase in entropy with increasing deformation strength with states corresponding to energies in the bulk of the spectrum obtaining values corresponding to the predictions of RMT. The entropy also shows a relatively smooth dependence on the energy. On the other hand, for the $H_{\text{QInt}}$ model the onset of chaos is slower, particularly for small $\epsilon$. Moreover the scaling of the critical coupling can be described by a power-law, $\epsilon_c \sim L^{-b}$, with an exponent intermediate between those expected for a generic gapless system $b=3$ and the weakly deformed XXZ model $b=2$. However, the scaling is not particularly clear due to the limited range of system lengths that are numerically available, and it can be also well described by an exponential form in both the gapped and gapless phases. Indeed we do not see a strong distinction between the two phases in the scaling exponent except for a modest decrease in exponent with increasing $J_z$. It may be the case that the quasi-integrable model is a partially delocalised perturbation in Fock space, and so generates an onset of chaos intermediate between a transition and crossover, or that it simply requires going to longer lengths to make a clear distinction. Related behaviour is seen in the energy eigenstates where the information entropy of states is generally lower than those of $H_{\text{dXYZ}}$ for small $\epsilon$ and additionally the variance is consistently larger so that the entropy does not become a smooth function of the energy. It would be interesting to explore whether it is possible to observe these distinctions in other physical quantities and in particular whether the quasi-integrable model thermalises slower than generic integrability breaking models i.e. on time-scales $\tau\propto \epsilon^{-b}$ with $b>2$. Similarly, it would be interesting to study matrix elements of few-body operators in these models to understand if there are deviations from the predictions of the ETH.  
	
	\section*{Acknowledgements}
	We would like to thank Bal\'azs Pozsgay and Federica Surace for useful comments.
	MdL was supported in part by SFI and the Royal Society for funding under
	grants UF160578, RGF$\backslash$ R1$\backslash$ 181011, RGF$\backslash$8EA$\backslash$180167 and RF$\backslash$
	ERE$\backslash$ 210373. MdL is also supported by ERC-2022-CoG - FAIM 101088193. YFA thanks FAPESP for financial support under grant \#2025/03661-5.
	
	\appendix
	
	\section{Second Order $\g$ coefficients} \label{app.g}
	Here we display explicitly the $\g$ coefficients for each integrable model at second order.
	
	\subsection{Model A}
	\begin{align*}
		\mathrm{g}\Delta_{\text{xy}}^{(2)} 
		&= -2 i \left(h_{\text{yz},+}^{(1)} \delta_{x,-}^{(1)} 
		+ h_{\text{xz},+}^{(1)} \delta_{y,-}^{(1)}\right); \\[0.15cm]
		\mathrm{gh}_{\text{xy},+}^{(2)} 
		&= 2 i \left(h_{\text{xz},+}^{(1)} \delta_{x,-}^{(1)} 
		- h_{\text{yz},+}^{(1)} \delta_{y,-}^{(1)}\right); \\[0.15cm]
		\mathrm{g}\delta_{x,+}^{(2)} 
		&= 2 \coth\!\left(\frac{\eta}{2}\right) 
		\mathrm{g}\delta_{z,+}^{(1)} h_{\text{xz},+}^{(1)} 
		- 2 i \delta_{y,-}^{(1)} \delta_{z,+}^{(1)}; \\[0.15cm]
		\mathrm{g}\delta_{y,+}^{(2)} 
		&= 2 \coth\!\left(\frac{\eta}{2}\right) 
		\mathrm{g}\delta_{z,+}^{(1)} h_{\text{yz},+}^{(1)} 
		+ 2 i \delta_{x,-}^{(1)} \delta_{z,+}^{(1)}; \\[0.15cm]
		\mathrm{gh}_{\text{yz},-}^{(2)} 
		&= -2 i \left(h_{\text{xy},-}^{(1)} \delta_{y,-}^{(1)} 
		+ \mathrm{csch}^2\!\left(\frac{\eta}{2}\right) 
		h_{\text{xz},+}^{(1)} \delta_{z,+}^{(1)}\right); \\[0.15cm]
		\mathrm{gh}_{\text{xz},-}^{(2)} 
		&= -2 i \left(h_{\text{xy},-}^{(1)} \delta_{x,-}^{(1)} 
		- \mathrm{csch}^2\!\left(\frac{\eta}{2}\right) 
		h_{\text{yz},+}^{(1)} \delta_{z,+}^{(1)}\right); \\[0.15cm]
		\mathrm{gh}_{\text{xy},-}^{(2)} 
		&= 2 \sinh(\eta)\, \mathrm{g}\sigma_{\text{xy}}^{(1)} 
		h_{\text{xy},-}^{(1)} 
		- 2 i \left(2 \sigma_{\text{xy}}^{(1)} \delta_{z,+}^{(1)} 
		+ \mathrm{csch}(\eta) \delta_{z,+}^{(2)}\right); \\[0.15cm]
		\mathrm{gh}_{\text{xz},+}^{(2)} 
		&= 2 \sinh(\eta)\, \mathrm{g}\sigma_{\text{xy}}^{(1)} 
		h_{\text{xz},+}^{(1)} \notag \\
		&\quad - i \left(2 \left(\Delta_z^{(1)} - \sigma_{\text{xy}}^{(1)}\right) 
		\delta_{y,-}^{(1)} 
		+ \tanh\!\left(\frac{\eta}{2}\right) 
		\delta_{y,-}^{(2)} -2 h_{\text{yz},+}^{(1)} \delta_{z,-}^{(1)} 
		+ \right);\\[0.15cm]
		\mathrm{gh}_{\text{yz},+}^{(2)} 
		&= 2 \sinh(\eta)\, \mathrm{g}\sigma_{\text{xy}}^{(1)} 
		h_{\text{yz},+}^{(1)} \notag \\
		&\quad + i \left( 2 \left(\Delta_z^{(1)} - \sigma_{\text{xy}}^{(1)}\right) 
		\delta_{x,-}^{(1)}+ \tanh\!\left(\frac{\eta}{2}\right) 
		\delta_{x,-}^{(2)} 
		-2 h_{\text{xz},+}^{(1)} \delta_{z,-}^{(1)} 
		\right); \\[0.15cm]
		\mathrm{g}\Delta_{z}^{(2)} 
		&= -2 i (\cosh(\eta) + 2)
		\left(h_{\text{yz},+}^{(1)} \delta_{x,-}^{(1)} 
		- h_{\text{xz},+}^{(1)} \delta_{y,-}^{(1)}\right) \notag \\
		&\quad - 4 i \cosh(\eta) 
		h_{\text{xy},-}^{(1)} \delta_{z,+}^{(1)}+ \cosh(\eta)\, \mathrm{g}\sigma_{\text{xy}}^{(2)} \notag \\
		&\quad + 2 \sinh(\eta)\, \mathrm{g}\sigma_{\text{xy}}^{(1)} 
		\left(\Delta_z^{(1)} 
		- \cosh(\eta)\sigma_{\text{xy}}^{(1)}\right).
	\end{align*}
	
	\subsection{Model B}
	\begin{align*}
		\mathrm{g}\delta_{x,+}^{(2)} &= 0, \quad \mathrm{g}\delta_{y,+}^{(2)} = 0, \\
		\mathrm{gh}_{\text{xz},-}^{(2)} &= 0, \quad 	\mathrm{gh}_{\text{yz},-}^{(2)} = 0.
	\end{align*}
	\begin{align*}
		\mathrm{gh}_{\text{xy},+}^{(2)}
		&= 2 \sinh(\eta)\, \mathrm{g}\sigma_{\text{xy}}^{(1)} 
		h_{\text{xy},+}^{(1)}
		- 2 i \left(
		- h_{\text{xz},+}^{(1)} \delta_{x,-}^{(1)}
		+ h_{\text{yz},+}^{(1)} \delta_{y,-}^{(1)}
		+ 2 \Delta_{\text{xy}}^{(1)} \delta_{z,-}^{(1)}
		\right); \\[0.15cm]
		\mathrm{gh}_{\text{xy},-}^{(2)}
		&= -2 i\, \mathrm{csch}(\eta)\, \delta_{z,+}^{(2)};\\[0.15cm]
		\mathrm{g}\Delta_{\text{xy}}^{(2)} 
		&= 2 \sinh(\eta)\, \mathrm{g}\sigma_{\text{xy}}^{(1)} 
		\Delta_{\text{xy}}^{(1)}
		- 2 i \left(
		h_{\text{yz},+}^{(1)} \delta_{x,-}^{(1)}
		- 2 h_{\text{xy},+}^{(1)} \delta_{z,-}^{(1)}
		+ h_{\text{xz},+}^{(1)} \delta_{y,-}^{(1)}
		\right) \\[0.15cm]
		\mathrm{g}\Delta_{z}^{(2)}
		&=  2 i (\cosh(\eta)+2)
		h_{\text{xz},+}^{(1)} \delta_{y,-}^{(1)}-2 i (\cosh(\eta)+2)
		h_{\text{yz},+}^{(1)} \delta_{x,-}^{(1)}
		\notag \\
		&\quad - 2 \sinh(\eta)\cosh(\eta)
		\mathrm{g}\sigma_{\text{xy}}^{(1)} \sigma_{\text{xy}}^{(1)}
		+ \cosh(\eta)\, \mathrm{g}\sigma_{\text{xy}}^{(2)} \notag \\
		&\quad + 2 \sinh(\eta)\, \mathrm{g}\sigma_{\text{xy}}^{(1)} 
		\Delta_{z}^{(1)}; \\[0.15cm]
		\mathrm{gh}_{\text{xz},+}^{(2)}
		&= 2 \sinh(\eta)\, \mathrm{g}\sigma_{\text{xy}}^{(1)} 
		h_{\text{xz},+}^{(1)} \notag \\
		&\quad - i \left(
		2 h_{\text{xy},+}^{(1)} \delta_{x,-}^{(1)}
		- 2 h_{\text{yz},+}^{(1)} \delta_{z,-}^{(1)}
		- 2 (\Delta_{\text{xy}}^{(1)} + \sigma_{\text{xy}}^{(1)} - \Delta_{z}^{(1)})
		\delta_{y,-}^{(1)} 
		+ \tanh\!\left(\frac{\eta}{2}\right)
		\delta_{y,-}^{(2)}
		\right); \\[0.15cm]
		\mathrm{gh}_{\text{yz},+}^{(2)}
		&= 2 \sinh(\eta)\, \mathrm{g}\sigma_{\text{xy}}^{(1)} 
		h_{\text{yz},+}^{(1)} \notag \\
		&\quad + i \left(
		2 h_{\text{xy},+}^{(1)} \delta_{y,-}^{(1)}
		- 2 h_{\text{xz},+}^{(1)} \delta_{z,-}^{(1)}
		+ 2 (\Delta_{\text{xy}}^{(1)} - \sigma_{\text{xy}}^{(1)} + \Delta_{z}^{(1)})
		\delta_{x,-}^{(1)} 	+ \tanh\!\left(\frac{\eta}{2}\right)
		\delta_{x,-}^{(2)} 
		\right).
	\end{align*}

	\section{Data Preparation}
	\label{app:data}
	\subsection*{Unfolding}
	\label{app:unfolding}
	In order to compare spectral statistics with RMT predictions it is useful to separate the spectrum of energy eigenvalues $\{E_1, E_2, \dots \}$ into a contribution that depends on a smoothed or averaged component and a fluctuation component which has a uniform average density. This is done by computing the cumulative number density
	\begin{align}
		n(E)=\int_{-\infty}^E \rho(x)dx =\sum_i \Theta(E-E_i)~.
	\end{align}
	and performing a local average. As a practical matter we perform this numerically by first sorting the energy eigenvalues, then numerically fitting a high-degree polynomial (we choose a degree 25 though the results are stable to variations in this parameter) to the data to define the smooth cumulative number density $n_{av}(E)$.
	In order to get a good fit to the data we must remove some data points from the edge of the spectrum, that is we ``clip" the spectrum. This clipping is further motivated by the physical behaviour of the edge states and is discussed in greater detail below.  The unfolded spectrum, $\{\epsilon_i\}$, is then defined by the averaged number density evaluated on each energy eigenvalue:
	\begin{align}
		\epsilon_i=n_{av}(E_i)
	\end{align}
	such that the unfolded spectrum has an approximately uniform density.

	\subsection*{Clipping}
	\label{app:clipping}
	One aspect of the data analysis is ``clipping" which refers to the removal of states from the ends of spectrum. The motivation arises from the fact that states near the edges do not behave in the same fashion as states from the middle of the spectrum. This can be seen in the entropy of the various energy eigenstates shown in Fig. \ref{fig:entropy_fits} but can also be seen it in the computation of the level statistics. In particular, in analysing the transition from Poisson to Wigner-Dyson we find different critical couplings depending on what proportion of the spectra we remove. 
	
	In Tab. \ref{tab:ham_rel_size} we consider the models $H_{\text{dXYZ}}$ and $H_{\text{QInt}}$ in the sector with $L=21$, momentum one and even magnetisation which has dimension 49,929. We compute the level-spacing distribution as we vary the deformation strength, $\epsilon$, for different proportions of clipping from $5\%$ to $45\%$ which we perform symmetrically at each end. By fitting to the Brody distribution we see that both the maximum value of the Brody parameter, $\omega_{\text{max}}$ and the critical coupling $\epsilon_c$ at which $\omega(\epsilon_c)=1/2$ depend on the clipping proportion. 
	\begin{table}[htbp]
		\centering
		\begin{tabular}{|l||cc||cc|}
			\hline
			&\multicolumn{2}{l||}{$H_{\text{dXYZ}}$}&\multicolumn{2}{l|}{$H_{\text{QInt}}$}\\
			\cline{1-5}
			Clip $\%$ &  $\omega_{\text{max}}$  &  $\epsilon_{c}$ & $\omega_{\text{max}}$  &  $\epsilon_{c}$ \\
			\hline\hline
			5  & 0.854  & 0.031 & 0.863 & 0.112\\
			10  & 0.883 &  0.029 & 0.889 & 0.109 \\
			15  & 0.937  &  0.026 & 0.959 &  0.104 \\
			\textbf{20} & \textbf{0.964} &  \textbf{0.025} & \textbf{0.986} & \textbf{0.101} \\
			25 & 0.984 &  0.025& 0.985 & 0.100 \\
			30 & 0.989 &  0.025 & 0.972 & 0.100 \\
			35 & 0.999 &  0.025 & 0.975 & 0.099\\
			40 & 0.996 &  0.025 & 0.988 & 0.098\\
			45 & 1.00 &  0.025 & 0.990 & 0.099\\
			\hline
		\end{tabular}
		\caption{Clipping effects for $L=21$, momentum-one, even magnetisation sectors.}
		\label{tab:ham_rel_size}
	\end{table} 
	In many regards it is to be expected that different states, and so different parts of the spectrum, become chaotic at different rates and in this regard there may not be a single critical coupling for the entire system. Perhaps an analogy can be found with classical systems where one sees certain regions of phase space become chaotic while in others invariant KAM torii will persist for finite values of the coupling. Nonetheless, one can see that for clippings around 15-20\% the results for the critical coupling become reasonably stable and as we wish to preserve as many states as possible, so that our numerical results are reliable even at relatively short lengths, we choose 20\% clippings. 
	
	\section*{Operator norms}
	\label{app:norms}
	\begin{figure}
		\vspace{0.3cm}
		\begin{center}
			\includegraphics[height=0.3\linewidth]{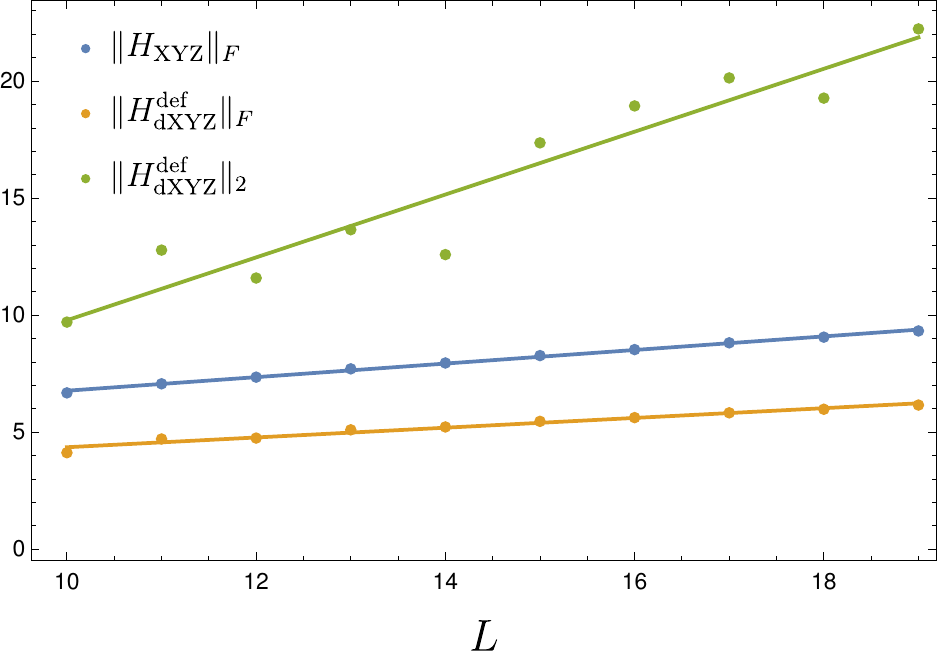}
			\hspace{3mm}
			\includegraphics[height=0.3\linewidth]{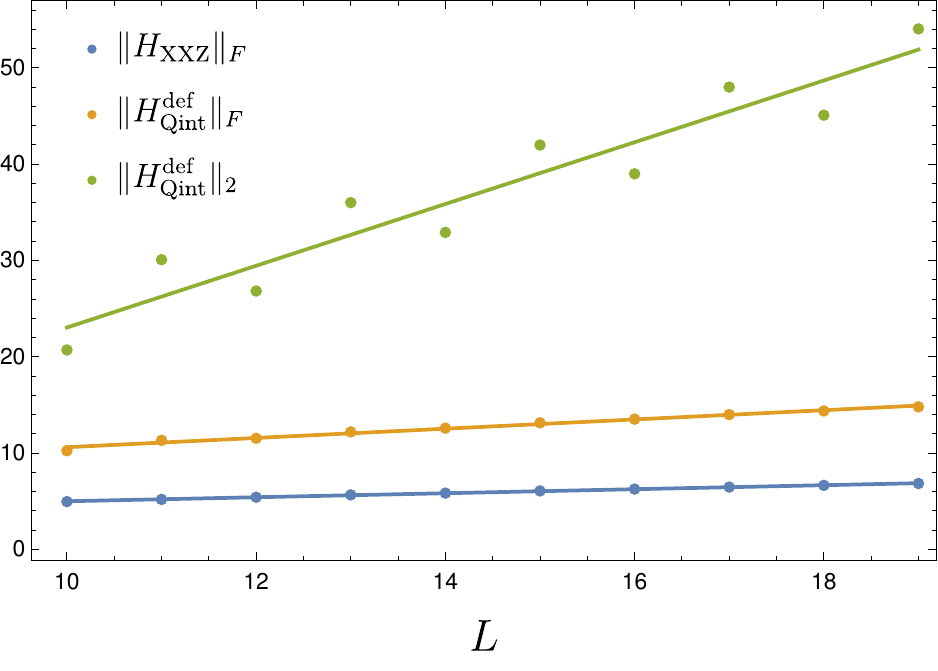}
		\end{center}
		\caption{(Left) Operator norms for the deformed XYZ model with $\Delta_{xy}=0.45$, $\sigma_{xy}=J_z=1.18$, $h_{xy, -}=\epsilon$ as a function of $L$ for states in the momentum-one, even magnetisation sector. (Right)  Operator norms for the QInt model with $\sigma_{xy}=1$, $J_z=0.68$, $\gamma=1$, $\alpha=0.82$, $\beta=1.43$, as a function of $L$ in the momentum-one, even-magnetisation sector.}	\label{fig:opnorms}
	\end{figure}
	In order to compare the strengths of deformations in different models it is useful to quantify the size of the deformation relative to the undeformed Hamiltonian. We can do this by computing the operator norm of the relative contributions. As each operator is a sum of the length of the spin-chain we expect these to grow linearly with $L$. A natural choice is the spectral norm
	\begin{align}
		\|\mathcal{O}\|_2=\text{sup}\{\|\mathcal{O}x\|:x\in V ~\& ~\|x\|=1\}
	\end{align}
	corresponding to the largest singular value of $\mathcal{O}$. Alternatively, and quicker to compute, is the normalised Frobenius norm
	\begin{align}
		\|\mathcal{O}\|_F=\sqrt{\frac{\Tr(\mathcal{O}^\dagger \mathcal{O})}{\Tr(\unit)}}~.
	\end{align}
	An advantage of the Frobenius norm is that it includes contributions from the entire spectrum and as we are interested in the bulk states this seems more appropriate. Finally, as can be seen in Fig. \ref{fig:opnorms}, it gives a relatively smooth result which can be extrapolated to higher lengths. 
	We numerically compute the norms for $L\in [10,19]$ in the momentum-one, even-magnetisation sectors for $H_{\text{dXYZ}}$ and $H_{\text{QInt}}$. We separately compute the Frobenius norms for the undeformed theories (resp. $H_{\text{XYZ}}$ and $H_{\text{XXZ}}$) and the deformations $H^{\text{def}}=H(\epsilon=1)-H(\epsilon=0)$. These can be seen to be clearly linear in $L$ and the corresponding fits are
	\begin{align}
		\|H_{\text{XYZ}}\|_F= 3.88 +0.29 L~, & ~~~\|H_{\text{XXZ}}\|_F= 2.91+0.21 L \nn\\
		\|H^{\text{def}}_{\text{dXYZ}}\|_F= 2.28+0.21 L ~,&~~~ \|H^{\text{def}}_{\text{QInt}}\|_F= 5.80+0.48 L
	\end{align}
	which we use to compute the norms for larger $L$.
	
	For comparison we compute the spectral norm for the deformations and here we see that while they are approximately linear there is some dependence on $L$. Given the norms we can then define an effective deformation strength
	\begin{align}
		\label{eq:eff_cpl}
		\epsilon_{\text{eff}}=\frac{\| H(\epsilon=1)-H(\epsilon=0)\|_F}{\|H(\epsilon=0)\|_F}\epsilon~.
	\end{align}
	However, for the most part we refrain from using the effective coupling as it doesn't appear to substantially alter our results; for example the volume scaling is modified only slightly, with, for example $b_{\text{eff}}=2.6$ compared to $2.63$ for the QInt model with $J_z=0.68$ which is within the standard error. 
	
	\addcontentsline{toc}{section}{References}
	\bibliography{NN}
	\bibliographystyle{utphys}

\end{document}